\documentclass[superscriptaddress,twocolumn,pre,showpacs,preprintnumbers]{revtex4-1}
\usepackage{amsmath,amssymb}
\usepackage{epsfig,graphics,color,calc,graphicx}
\usepackage{mathrsfs}

\newcommand{\rr}[1]{{\normalfont\textrm{#1}}}

\newcommand{\drift}{\delta}

\begin{document}
\title{A lattice model of reduced jamming by barrier}

\author{Emilio N.M.\ Cirillo}
\email{emilio.cirillo@uniroma1.it}
\affiliation{Dipartimento di Scienze di Base e Applicate per
             l'Ingegneria, Sapienza Universit\`a di Roma,
             via A.\ Scarpa 16, I--00161, Roma, Italy.}

\author{Oleh Krehel}
\email{o.krehel@tue.nl}
\affiliation{ICMS -- Institute of Complex Molecular Systems,
Department of Mathematics and Computer Science,
  Eindhoven University of Technology,
  P.O.\ Box 513, 5600 MB Eindhoven, The Netherlands.}

\author{Adrian Muntean}
\email{adrian.muntean@kau.se}
\affiliation{Department of Mathematics and Computer Science,
Karlstad University, Sweden.}

\author{Rutger van Santen}
\email{R.A.v.Santen@tue.nl}
\affiliation{ICMS -- Institute for Complex Molecular Systems,
Faculty of Chemical Engineering,
  Eindhoven University of Technology,
  P.O.\ Box 513, 5600 MB Eindhoven, The Netherlands,}


\begin{abstract}
We study an asymmetric simple exclusion process in a strip in the
presence of a solid impenetrable barrier. We focus on the effect of the
barrier on the residence time of the particles, namely,
the typical time needed by the particles to cross the whole strip.
We explore the conditions
for reduced jamming when varying the environment (different drifts,
reservoir densities, horizontal diffusion walks,
etc.). Particularly, we discover an interesting non--monotonic
behavior of the residence time as a function of the barrier
length. Besides recovering by means of both the lattice dynamics and
mean--field model well--known aspects like faster--is--slower effect and
the intermittence of the flow, we propose also a birth--and--death
process and a reduced one--dimensional model with variable barrier
permeability to capture qualitatively the behavior of the
residence time with respect to the parameters. We report our first steps
towards the understanding to which extent the presence of obstacles
can fluidize pedestrian and biological transport in crowded
heterogeneous environments.
\end{abstract}





\maketitle


\section{Introduction}
\label{Introduction}
Lattice models of particle flow may show surprisingly rich behavior
even when only exclusion of a particle on the same
site
is considered
\cite{Rutger}. Complex percolation behavior arises in particular at
increased particle concentration (see \cite{AS} for a modern account
on percolation theory,  \cite{Padding} for a case study related to the motion of colloids in narrow channels, and \cite{Transport} for percolation effects in
transportation in more general complex systems).  In this paper, we introduce a two
dimensional asymmetric simple exclusion random walk model with
diffusion and drift. The model aims at capturing the effect of the barrier
positioned in the strip on the corresponding residence times,
i.e., the time needed by a particle to cross the strip.

More precisely,
we consider a (say) vertical strip and measure the time that a particle
entering the strip at the top side takes to exit the strip through the
bottom side, under the assumption that the three other boundaries
act as reflecting boundaries.
This typical time will be called \emph{residence} time.

We find an interesting
non--linear dependence on the length of this barrier  when
simulating the evolution of a high particle density in the strip. Instead of the
expected increase in the residence time, at particular conditions we surprisingly notice a
decrease in residence times with increasing barrier length. This
reminds us of the Braess paradox, discovered when traffic flow
unexpectedly decreases, whereas an inhibitive traffic access barrier
is removed (cf.\ \cite{Braess}). This confirms once more the fact that
as population densities and the number of interactions between particles
(agents, people, financial stocks, etc.) increase, so does the probability
of emergent phenomena.

Our modeling approach and simulation results are
potentially useful when trying to forecast the motion of pedestrian flows in open (heterogeneous) spaces. It has for instance been found
that flocking of sheep \cite{Garcimartin,Zuriguel} is helped by
introducing a barrier before an exit point.  Also high density
particle flow through an orifice that leads to jamming has been found
to have less jamming when a barrier is put in front of the orifice (see, for instance, \cite{Buzna} and \cite{Tomoeda} for crowd dynamics scenarios when the flow is improved by the presence of an obstacle in front of the exit).

\begin{figure*}[t]
  \centering
  \begin{tabular}{ll}
    \includegraphics[width=0.45\textwidth]{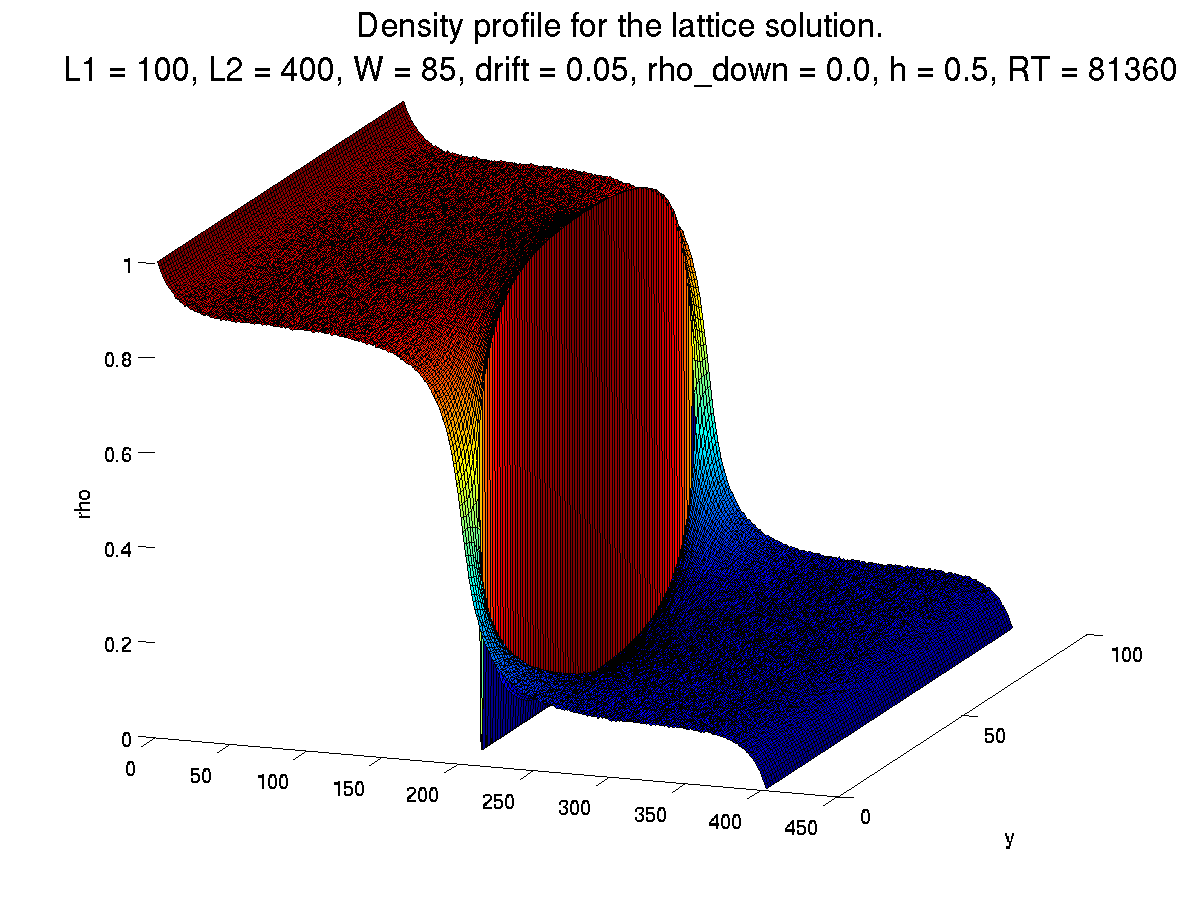}
    &
      \includegraphics[width=0.45\textwidth]{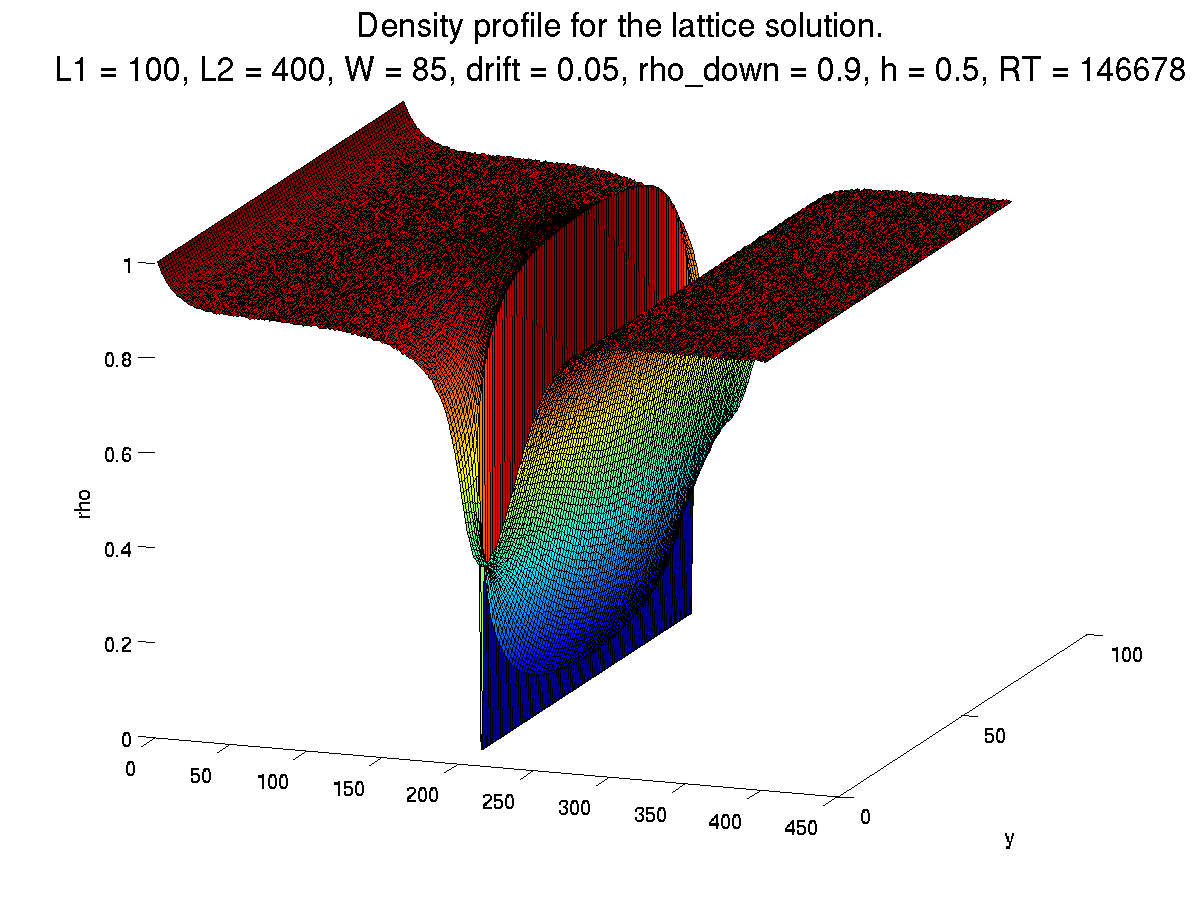}\\
    \includegraphics[width=0.45\textwidth]{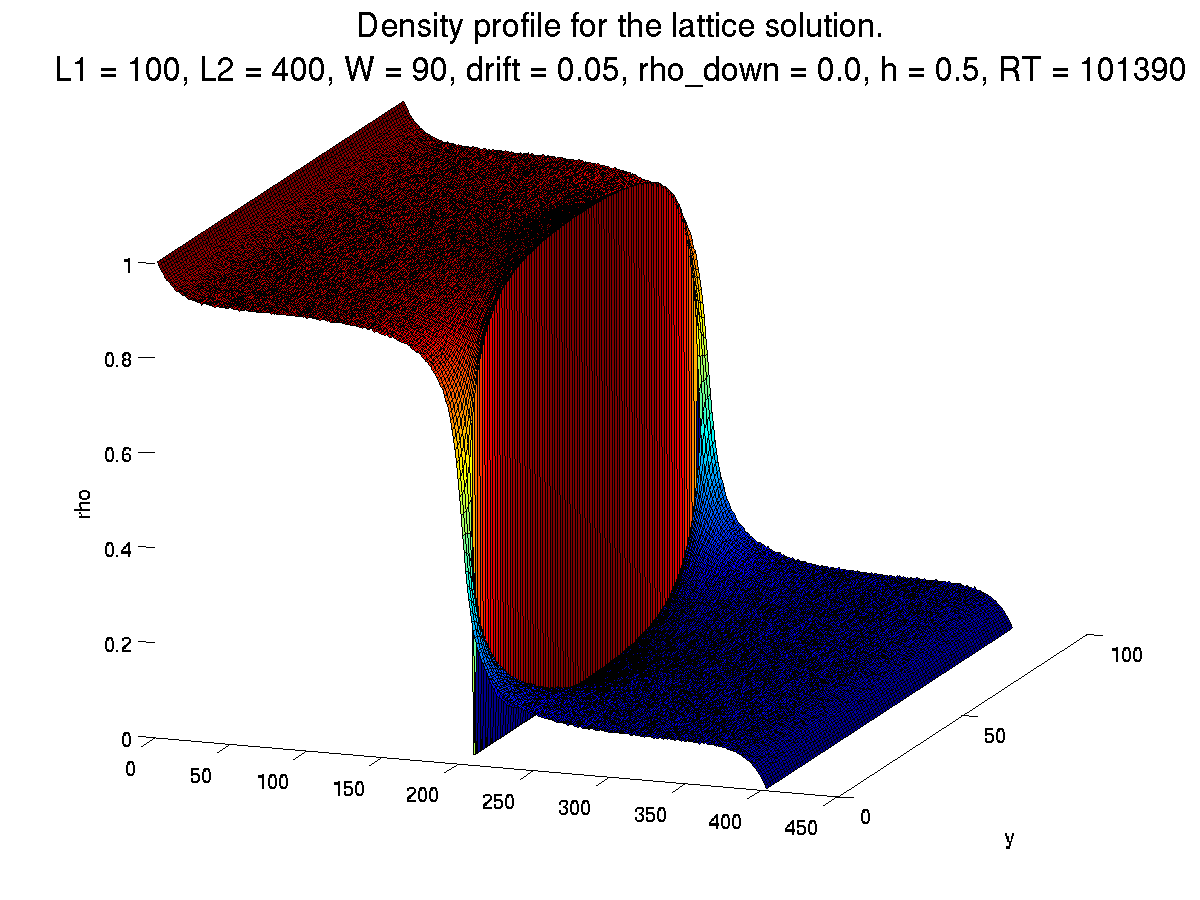}
    &
      \includegraphics[width=0.45\textwidth]{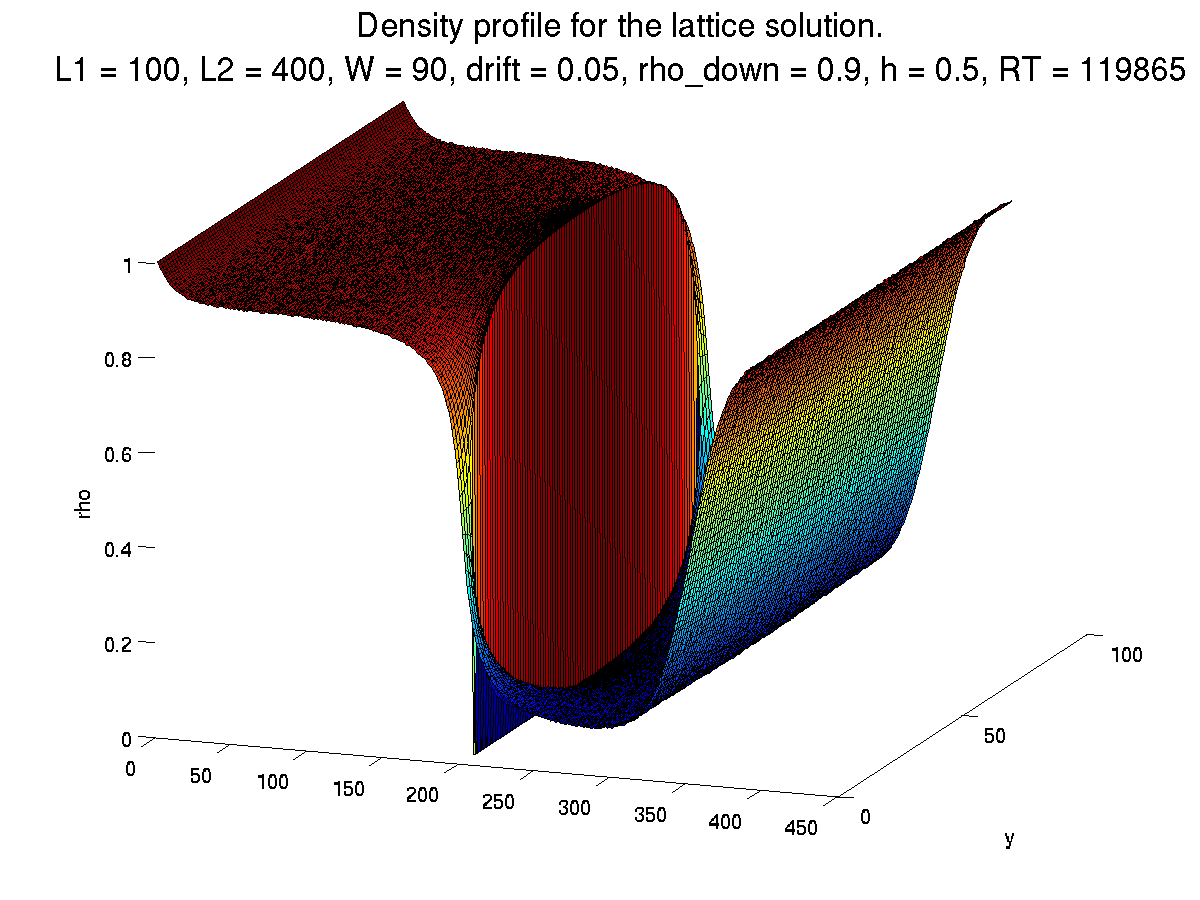}\\
    \includegraphics[width=0.45\textwidth]{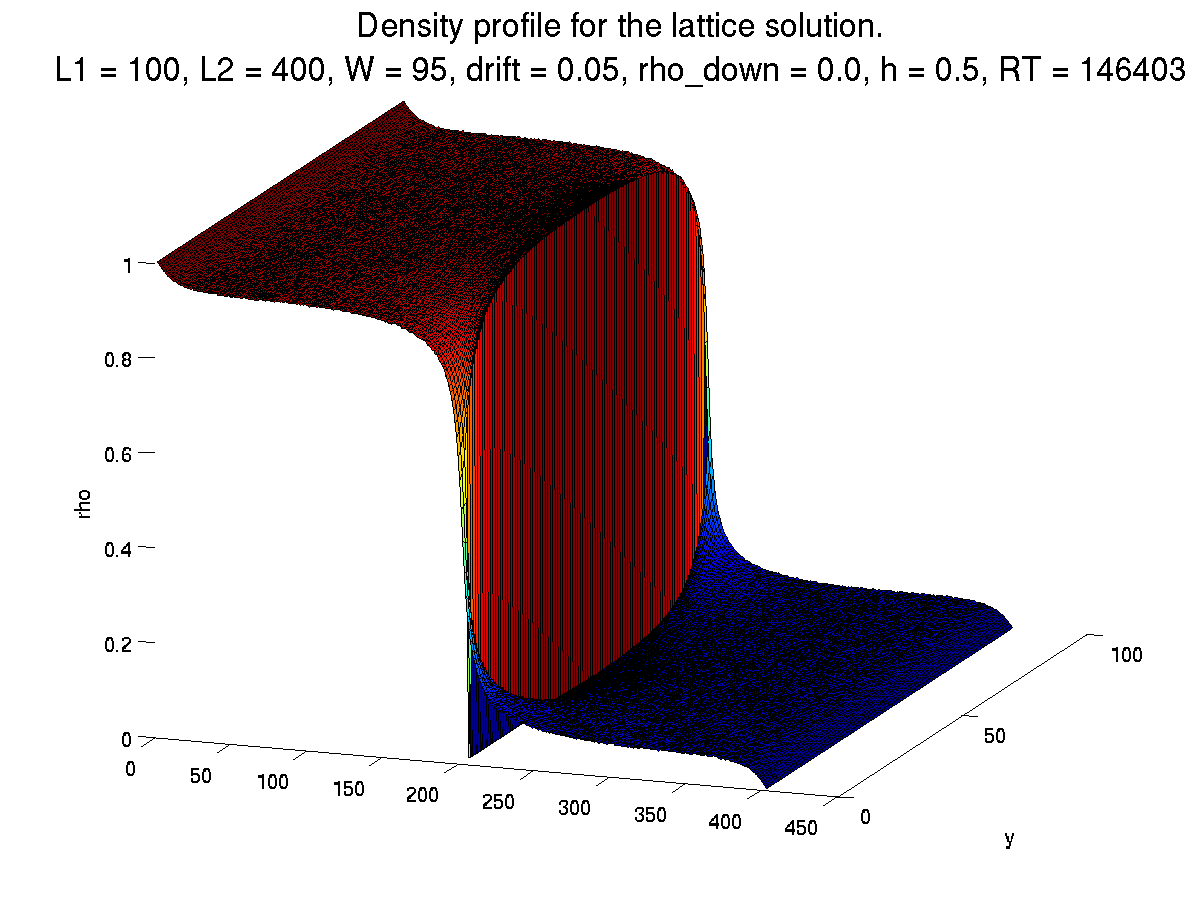}
    &
      \includegraphics[width=0.45\textwidth]{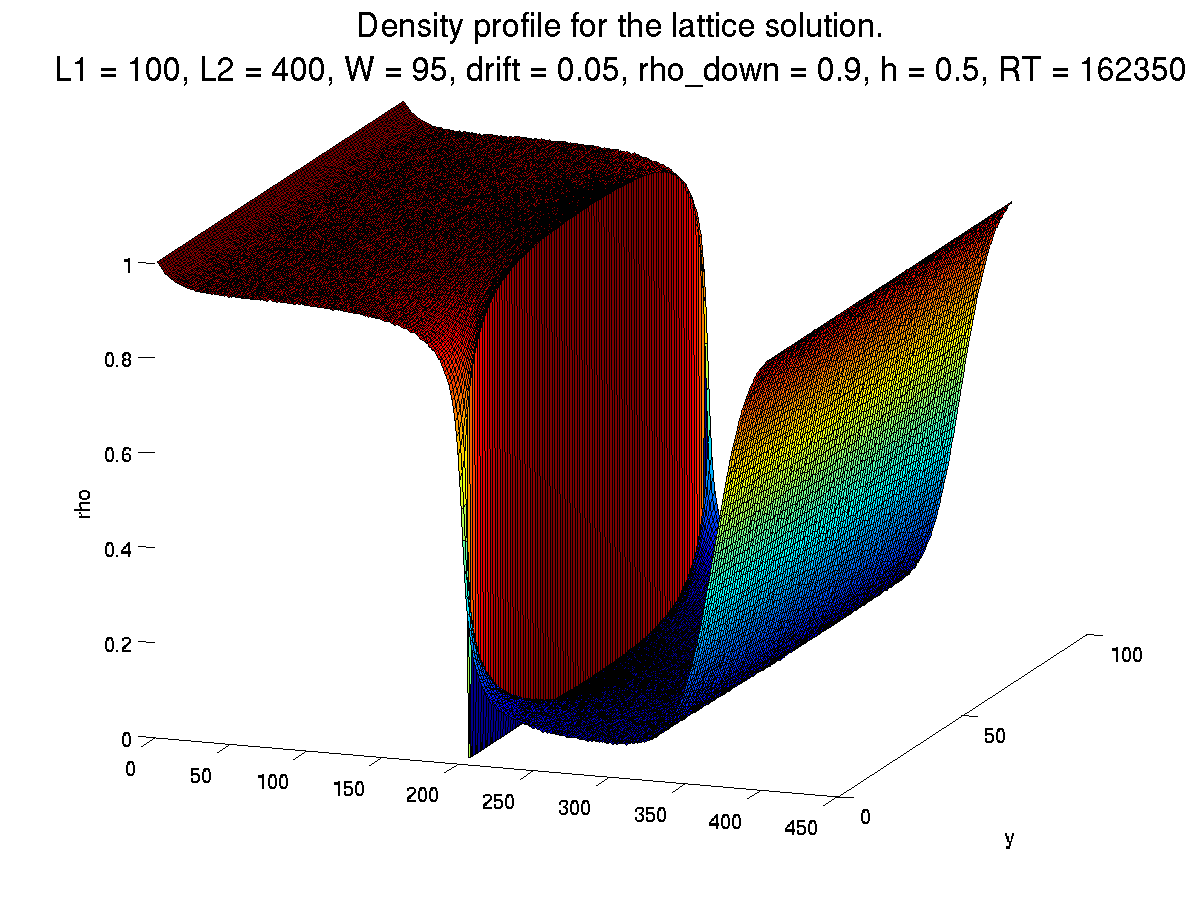}
  \end{tabular}
  \caption{2D density profiles. On the left, $\rho_\rr{d}=0.0$ and the
    average residence times are $81359.8, 101390, 146403$.  On the
    right, $\rho_\rr{d}=0.9$ and the average residence times are
    $146678, 119865, 162350$. The other parameters are
    $L_1=100$, $L_2=400$, $h=0.5$, $\delta=0.05$, $\rho_\rr{u}=1$,
    $O_2=3$, and $W=85$ (top), $W=90$ (middle), $W=95$ (bottom).}
  \label{fig:2D-density-drift-nonzero}
\end{figure*}

We have explored extensively in a previous paper (see \cite{Rutger})
the two dimensional diffusion--drift strip lattice model used in this
context, but without barriers.  On the two dimensional lattice a
discrete stochastic process is simulated controlled by top and bottom
reservoir densities. The displacement probabilities of the particles
are in four directions: $h$, $u$, $d$, with $h+u+d=1$.  Displacements
only can occur to a square lattice site that is unoccupied.
The horizontal displacement probability perpendicular to the flow
direction of the strip
is $h/2$, whereas $u$ and $d$ are the upward and downward
displacement probabilities. The model describes diffusion as well nonlinear
convection when $d-u$ is different from zero.

When the drift (pointing out
in the top-down direction)
is non--zero, our stochastic simulations show a phase
transition in the dependence of simulated average particle residence
time as a function of the barrier length $W$. This phase transition is
only found when the density
$\rho_\rr{d}$, i.e., the bottom reservoir density,
exceeds a particular value, while the range
of barrier lengths of decreases in residence time depends on the choice of the drift
value. In the absence of the drift, alike phase transitions do not
happen (as predicted for instance in \cite{Krug} and references cited
therein).

Denote our vertical strip by $\Omega$
and refer to the internal obstacle as  $\mathcal{O}$, see
Figure \ref{f:reticolo} for a sketch of the geometry.
The physical basis of this phenomenon can be understood based on the
particle density profiles. The calculated density profiles around the
phase transition are shown for a particular situation in
Figure~\ref{fig:2D-density-drift-nonzero} (the meaning of the
parameters listed in the caption will be explained
in Section~\ref{Method}). One notes the transformation of
a convex--to--concave density profile behind the barrier when the
barrier width is moved through the phase transition regime.  This
density profile can be well approximated as solution to the mean--field
equation
\begin{equation}
\label{MF}
\frac{\partial \rho}{\partial t}
=
\frac{h}{2}\frac{\partial^2\rho}{\partial y^2}
+\frac{1-h}{2}\frac{\partial^2\rho}{\partial x^2}
-\delta(1-h)\frac{\partial}{\partial x}(\rho(1-\rho))
\end{equation}
in $\Omega\setminus\mathcal{O}$,
endowed with the initial condition
\begin{equation}
\label{MF01}
\rho(0,y,x)=0  \mbox{ in } \Omega\setminus\mathcal{O}
\end{equation}
and the boundary conditions
\begin{equation}
\label{MF02}
\rho(t,y,0)=\rho_\rr{u}, \,\, \rho(t,y,L_2)=\rho_\rr{d},
\end{equation}
and
\begin{equation}
\label{MF03}
\frac{\partial\rho (t,0,x)}{\partial y}
=\frac{\partial{\rho (t,L_1,x)}}{\partial y}
=\nabla \rho\cdot n_{\partial \mathcal{O}}=0.
\end{equation}
Here $n_{\partial \mathcal{O}}$ denotes the outer normal along
the boundary of the obstacle $\mathcal{O}$.

It occurs to us that there may
 not be  too much dependence in the density profile on the $y$ variable
and we can approximate the two dimensional density profile with its
one dimensional counterpart $\tilde\rho(x)$
that we obtain by integrating out the $y$
variable.
This one dimensional density profile can be then
used to calculate the residence time estimate that is given from the
mean field expression
\begin{equation}
\label{rt-mf}
  R= - \frac{2}{(1-h)\partial_x\tilde \rho(0)}
\int_0^{L_2}\tilde\rho(x)\, dx.
\end{equation}
This expression \cite[equation~(5.35)]{Rutger}
shows that the average particle residence time is
determined by the derivative of the density at the entrance of the
strip and the integrated density. The convex to concave density
profile change behind the barrier
in Figure~\ref{fig:2D-density-drift-nonzero} indicates a large change
in the particle density, that, as we will see, is responsible to a
significant extend to the phase transition behavior.  We have
discussed previously in \cite{Rutger} that the mean field equation
(\ref{MF}) is only valid in a limited regime of the parameter space,
there a birth--and--death random walk model providing an alternative
approach to calculate the residence time is proposed.

Non--linear behaviors in the residence time are not limited to the
dependence on the barrier width (that occurs when the drift is not
zero). Parametric dependencies can turn to be non--monotonic as
well. It is worth noting that in absence of drift, the
dependence on the barrier width always turns into a monotonic decrease
of the residence time with increasing width. However, mind that this
decrease does not uniformly scale with the lateral strip dimension.
When the residence times are considered at similar ratios of barrier
width and strip lateral dimension, the corresponding residence times
are found to increase with increasing the strip lateral dimension. The
effect depends on the horizontal hopping probability and diminishes
when the hopping frequency becomes larger.

In addition to the numerical solution of the residence time on the one
dimensional density profiles determined by averaging the density of
the two dimensional simulations, approximate analytical solutions are sought for
the corresponding viscous one-dimensional Burgers equation, which
has then to be solved together with the proper boundary conditions.

The paper is organized as follows. In Section \ref{Method}, we introduce the lattice model and the
different methods to approach the barrier problem. This is to be
followed by the presentation of our results in Section
\ref{Results} and \ref{Results-nz}.
Essentially, we compare the two dimensional model
simulations with the output of the 1D model and give
evidence of the occurrence of a phase transition in one
dimension. The paper is concluded with a short discussion of the
results in Section \ref{Discussion}.

\section{Models and methods}
\label{Method}
In this section, we introduce the models we plan to study to address
the problem discussed in the introduction and we shall also give a
brief account of our main methods.

\subsection{Lattice dynamics}\label{lattice}
The lattice model we discuss in this paper
is the same as the one introduced in \cite{Rutger}, excepting
for the presence
of the obstacle. Nevertheless,
for the sake of clarity we define the model in detail.

Take $L_1,L_2\in\mathbb{N}$. Let $\Lambda\subset\mathbb{Z}^2$
denote the \emph{strip} $\{1,\dots,L_1\}\times\{1,\dots,L_2\}$.
We say that the coordinate directions $1$ and $2$ of the
strip are respectively the \emph{horizontal} and the \emph{vertical}
direction. We  accordingly use the words \emph{top},
\emph{bottom}, \emph{left}, and \emph{right}.
On $\Lambda$ we define a discrete time stochastic process controlled by the
parameters
$\varrho_\rr{u},\varrho_\rr{d}\in[0,1]$
and
$h,u,d\in[0,1]$
such that
$h+u+d=1$.
The meaning of the parameters is clarified in what follows.

\setlength{\unitlength}{0.8pt}
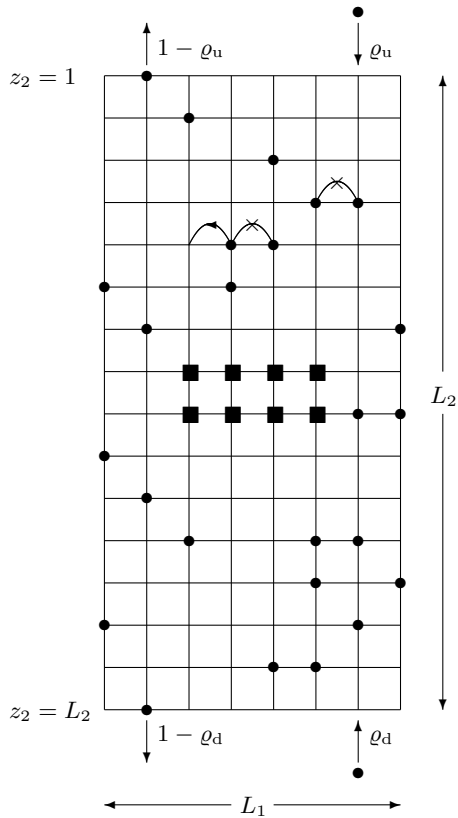
\begin{figure}[h]
\begin{picture}(400,330)(-40,20)
\thinlines
\multiput(40,0)(20,0){8}{\line(0,1){300}}
\multiput(40,0)(0,20){16}{\line(1,0){140}}
\put(100,-45){\vector(-1,0){60}}
\put(120,-45){\vector(1,0){60}}
\put(104,-49){${L_1}$}
\put(200,140){\vector(0,-1){140}}
\put(200,160){\vector(0,1){140}}
\put(-5,296){${z_2=1}$}
\put(-5,-4){${z_2=L_2}$}
\put(194,145){${L_2}$}
\put(76.5,135.5){$\blacksquare$}
\put(96.5,135.5){$\blacksquare$}
\put(116.5,135.5){$\blacksquare$}
\put(136.5,135.5){$\blacksquare$}
\put(76.5,155.5){$\blacksquare$}
\put(96.5,155.5){$\blacksquare$}
\put(116.5,155.5){$\blacksquare$}
\put(136.5,155.5){$\blacksquare$}
\put(40,40){\circle*{5}}
\put(40,120){\circle*{5}}
\put(40,200){\circle*{5}}
\put(60,0){\circle*{5}}
\put(60,100){\circle*{5}}
\put(60,180){\circle*{5}}
\put(60,300){\circle*{5}}
\put(80,80){\circle*{5}}
\put(80,280){\circle*{5}}
\put(100,200){\circle*{5}}
\put(100,220){\circle*{5}}
\put(120,20){\circle*{5}}
\put(120,220){\circle*{5}}
\put(120,260){\circle*{5}}
\put(140,20){\circle*{5}}
\put(140,60){\circle*{5}}
\put(140,80){\circle*{5}}
\put(140,240){\circle*{5}}
\put(160,40){\circle*{5}}
\put(160,80){\circle*{5}}
\put(160,140){\circle*{5}}
\put(160,240){\circle*{5}}
\put(180,60){\circle*{5}}
\put(180,140){\circle*{5}}
\put(180,180){\circle*{5}}
\thinlines
\put(60,305){\vector(0,1){20}}
\put(65,310){${1-\varrho_\textrm{u}}$}
\put(160,325){\vector(0,-1){20}}
\put(165,310){${\varrho_\textrm{u}}$}
\put(160,330){\circle*{5}}
\put(60,-5){\vector(0,-1){20}}
\put(65,-15){${1-\varrho_\textrm{d}}$}
\put(160,-25){\vector(0,1){20}}
\put(165,-15){${\varrho_\textrm{d}}$}
\put(160,-30){\circle*{5}}
\qbezier(100,220)(110,240)(120,220)
\put(105.5,226.5){${\times}$}
\qbezier(80,220)(90,240)(100,220)
\put(93,229.5){\vector(-1,0){5}}
\qbezier(140,240)(150,260)(160,240)
\put(145.5,246.5){${\times}$}
\end{picture}
\vskip 2. cm
\caption{Schematic representation of the model
in the presence of the barrier.
Solid squares represent the particles at rest modeling the
obstacle.}
\label{f:reticolo}
\end{figure}

The \emph{configuration} of the system at time $t\in\mathbb{Z}_+$
is given by the positive integer $n(t)$ denoting
the number of particles in the system at time $t$ and by the two
collections of integers
$x_1(1,t),\dots,x_1(n(t),t)\in\{1,\dots,L_1\}$ and
$x_2(1,t),\dots,x_2(n(t),t)\in\{1,\dots,L_2\}$ denoting, respectively,
the horizontal and the vertical coordinates of the $n(t)$ particles
in the strip $\Lambda$ at time $t$.
The $i$--th particle, with $i=1,\dots,n(t)$,
is then associated with the site $(x_1(i,t),x_2(i,t))\in\Lambda$ which
is called \emph{position} of the particle at time $t$.
A site associated with a particle a time $t$ will be said to be
\emph{occupied} at time $t$, otherwise we shall say that it is
\emph{free} or \emph{empty} at time $t$.
Fix $n(0)=0$.

At each time $t\ge1$ we first set
$n(t)=n(t-1)$ and then
repeat the following algorithm $n(t-1)$ times.
Essentially, at each step of the dynamics, a number of particles equal to the number of particles in the system at the end of the preceding time $n(t-1)$ is tentatively moved.
One of the three actions
\emph{insert a particle through the top boundary},
\emph{insert a particle through the bottom boundary},
and
\emph{move a particle in the bulk} is performed with
the corresponding probabilities
$\varrho_\rr{u}L_1/
(\varrho_\rr{u}L_1+\varrho_\rr{d}L_1+n(t))$,
$\varrho_\rr{d}L_1/
(\varrho_\rr{u}L_1+\varrho_\rr{d}L_1+n(t))$,
and
$n(t)/
(\varrho_\rr{u}L_1+\varrho_\rr{d}L_1+n(t))$.

\emph{Insert a particle through the top boundary.\/}
Chose at random with uniform probability the integer $i\in\{1,\dots,L_1\}$
and, if the site $(1,i)$ is empty, with probability $d$
set $n(t)=n(t)+1$ and add a particle to site $(1,i)$.

\emph{Insert a particle through the bottom boundary.\/}
Chose at random with uniform probability the integer $i\in\{1,\dots,L_1\}$
and, if the site $(L_2,i)$ is empty, with probability $u$
set $n(t)=n(t)+1$ and add a particle to site $(L_2,i)$.

\emph{Move a particle in the bulk.\/}
Chose at random with uniform probability one of the $n(t)$ particles
in the bulk.
The chosen particle is moved according to the following rule:
one of the four neighboring sites of the one occupied by the particle is chosen
at random with probability $h/2$ (left), $u$ (up), $h/2$ (right), and
$d$ (down). If the chosen site is in the strip (not on
the boundary) and it is free, the particle is moved there
leaving empty the site occupied at time $t$.
If the chosen site is on the boundary of the strip the dynamics is
defined as follows:
the left boundary $\{(0,z_2),\;z_2=1,\dots,L_2\}$ and
the right boundary $\{(L_1+1,z_2),\;z_2=1,\dots,L_2\}$
are \emph{reflecting} (homogeneous Neumann boundary conditions)
in the sense that a particle trying to jump there is not moved.
The bottom and the top
boundary conditions are stochastic in the sense
that when a particle tries to jump to a site $(z_1,0)$, with
$z_1=1,\dots,L_1$, such a site has to be considered occupied
with probability $\varrho_\rr{u}$ and free with probability
$1-\varrho_\rr{u}$,
whereas
when a particle tries to jump to a site $(z_1,L_2+1)$, with
$z_1=1,\dots,L_1$, such a site has to be considered occupied
with probability $\varrho_\rr{d}$ and free with probability
$1-\varrho_\rr{d}$.
If the arrival site is considered free
the particle trying to jump there is removed
by the strip $\Lambda$ (it is said to \emph{exit} the system)
and the number of particles is reduced by one, namely, $n(t)=n(t)-1$.
If the arrival site is occupied the particle is not moved.

\emph{Particle meets barrier.\/}
The impenetrable barrier is modeled by a rectangular region
of width $W$ and height $O_2$ which is constantly occupied
by particles at rest.
Hence,
particles moving on the lattice must do back step and/or
lateral jump this region.

It is worth noting that the model is a Markov
chain $\omega_0,\omega_1,\dots,\omega_t,\dots$
on the
\emph{state} or \emph{configuration space}
$\Omega:=\{0,1\}^\Lambda$ with transition probability
that can be deduced by the algorithmic definition.

This model will be studied via Monte Carlo simulations.
We will let evolve the process for a time (\emph{termalization time})
sufficiently long until the system reach the state. After that,
we shall measure the two--dimensional density profile by averaging
the occupation number at each site of the lattice (see,
for instance, Figure~\ref{fig:2D-density-drift-nonzero}).

Moreover, we shall also measure the residence time by averaging, at
stationarity, the time needed by a particle entered through the
top boundary to exit through the bottom one.
In this computation the top boundary condition will be chosen
to be $\rho_\rr{u}=1$ so that particles will not be allowed
to leave the system
through the top boundary.

In the study of the residence time we shall find two very different
pictures in the case in which the dynamics will
be either biased or not along the vertical direction.
A special role, hence, will be played by the parameter
\begin{equation}
\label{drift}
\delta=\frac{d-u}{d+u}
\end{equation}
which will be called \emph{drift}.

For more details we refer the reader to \cite{Rutger} where a complete
account on these techniques has been provided.

\subsection{Mean field dynamics}
The mean field equation \eqref{MF} corresponds to the
lattice dynamics presented in Subsection \ref{lattice}. It
is derived in full details in \cite{Rutger},
using arguments very much inspired from \cite{Landman}.
We refer the reader to these papers for the details of the
derivation of the mean field model and particularly to \cite{Rutger}
for a detailed investigation of its validity range
depending on the relative sizes of the most influential model parameters.
The novelty here is the presence of the obstacle.
The derivations follow similarly under the assumption that the obstacle
is impenetrable.

This mean field model is studied via a finite element approach.
The problem \eqref{MF}--\eqref{MF03} is integrated numerically and
the density profile $\rho(y,x)$ is found. Then the residence
time is computed by means of the equation \eqref{rt-mf}.

We used the Finite Element Numerics toolbox DUNE
\cite{bastian2008generic} to implement a solver for the model.  We
used quadratic Lagrange elements and the Newton method to deal with
the nonlinear drift term.

\subsection{One--dimensional reduction}
\label{onedim}
We propose a twofold reduction of the Mean Field model.
This way, we reduce the dimensionality of the model from 2D to 1D and compensate,
based on an effective transport coefficient,
for the presence of the obstacle.
For this we use a porous media modeling approach where parameters
like obstacle porosity and tortuosity will be used in the 1D context.
Similar arguments are indicated, for instance, in \cite{Bear}.

It occurs to us that there may
be not too much dependence in the density profile on the $y$ variable
and we can approximate the two dimensional density profile with its
one dimensional counterpart that we obtain by integrating out the $y$
variable. After integration, the $x$ coordinates that correspond to
the place where the obstacle was in two dimensions, are designated to
have a smaller diffusion coefficient to account for that obstacle.

In our initial approximation, we consider the diffusion
coefficient and the drift to be porosity and tortuosity based
via the coefficient
\begin{align}
  \label{eq:diffusion-coeff}
  \lambda(x)=
  \begin{cases}
    {\displaystyle F(h)\frac{L_1-W}{L_1}}
      & x\in[\frac{L_2-O_2}{2},\frac{L_2+O_2}{2}]\\
    1         & \text{otherwise.}
  \end{cases}
\end{align}
For convenience we also let
$\alpha:=F(h)(L_1-W)/L_1$.
Here, the ratio  $(L_1-W)/L_1$ is the porosity, while $F(h)$ is the currently unknown function of the horizontal
displacement probability $h$.  This plays the role of the tortuosity. It is expected that $F(h)\in (0,1)$.
In this very basic approximation porosity and tortuosity effects
are independent (multiplicative), so that the no obstacle
case is recovered for
$W=0$ and $F(h)=1$ in the expression (\ref{eq:diffusion-coeff}).
An increase in $W$ results in a decrease
in $\lambda(x)$ in the region $x\in[(L_2-O_2)/2,(L_2+O_2)/2]$,
which is also the expected behavior from the lattice model.

The 1D Mean Field equation reads
\begin{equation}
\label{eq:mean-field-1d}
\frac{d}{dx}\Big[\lambda(x)\Big(
\frac{1}{2}\frac{d\rho}{dx}-\drift\frac{d}{dx}(\rho(1-\rho))\Big)\Big]=0
\end{equation}
with the boundary conditions
\begin{equation}
\label{eq:mean-field-1d-bc}
\rho(0)=1
\;\;\textrm{ and }\;\;
\rho(L_2)=\rho_\rr{d}.
\end{equation}

On the basis of the density profile obtained by solving
\eqref{eq:mean-field-1d}, it is possible to compute the residence time
via a standard argument, see, e.g., \cite[Section~5.6]{Rutger}. We find
\begin{equation}
  \label{eq:residence-time}
  R=-\frac{2}{\rho'(0)}\int_{0}^{L_2}\rho(x),
\end{equation}
which is the analogous of equation \eqref{rt-mf}.

We will see in the next Section that the reduced model
(\ref{eq:mean-field-1d}) and (\ref{eq:mean-field-1d-bc})
is a convenient approximation of the 2D mean--field model with obstacle
in the zero drift case.
In this context the model will solved explicitly and the
density profile will be computed. Then
 we will compute the residence time using again \eqref{rt-mf}.

\section{Zero Drift Case}
\label{Results}
We consider the lattice model introduced in Section~\ref{lattice}
on the lattice strip of size $L_1\times L_2$ in absence of drift,
namely, for $\delta=0$.
Our simulations will be run mainly for
$L_1=100$, $L_2=400$, $h=0.5$, $\rho_\rr{u}=1$, and $\rho_\rr{d}=0,0.9$.
But in some cases we shall also consider the values $L_1=200$ and
$h=0.3,0.4$.
Our obstacle is of size $W\times O_2$ and is placed in the middle
of the strip. The typical values used in the simulations for the width
$W$ of the obstacle are $10,20,...,90$.
Its height $O_2$ will always be equal to $3$.

In this case, since particles do not experience any
external drift, we expect that the stationary density
profile will poorly depend on the horizontal lattice coordinate.
For this reasons it appears reasonable to compare our Monte Carlo
results for the lattice model with estimates based on the
one dimensional model introduced in Section~\ref{onedim}.

\subsection{Solution to the 1D model}
\label{1dreduction}
For $\delta=0$ the model in Section~\ref{onedim} simplifies
and a thorough analytical treatment is possible.
The 1D equation \eqref{eq:mean-field-1d}
is a linear diffusion equation with a piecewise
constant diffusion coefficient. Its solution is
piecewise linear on intervals $[0,(L_2-O_2)/2]$,
$[(L_2-O_2)/2,(L_2+O_2)/2]$, $[(L_2+O_2)/2,L_2]$,
and we can express it in the form
\begin{equation}
  \label{eq:fem-approx}
  \rho(x)=\rho_\rr{u}T_0(x)+aT_1(x)+bT_2(x)+\rho_\rr{d}T_3(x),
\end{equation}
where the coefficients $a$ and $b$ are the unknowns. The functions
$T_i$ are the linear pyramid functions. Their derivatives are
$T_0'(x)=-2/(L_2-O_2)$ on $[0,(L_2-O_2)/2]$ and $0$ otherwise,
\begin{align}
  &T_1'(x)=
    \begin{cases}
      \frac{2}{L_2-O_2}&\text{on }[0,(L_2-O_2)/2],\\
      -\frac{1}{O_2}&\text{on }[(L_2-O_2)/2,(L_2+O_2)/2],\\
      0&\text{otherwise}\nonumber
    \end{cases}
\end{align}
\begin{align}
  &T_2'(x)=
    \begin{cases}
      \frac{1}{O_2}&\text{on }[(L_2-O_2)/2,(L_2+O_2)/2],\\
      -\frac{2}{L_2-O_2}&\text{on }[(L_2+O_2)/2,L_2],\\
      0&\text{otherwise.}\nonumber
    \end{cases}
\end{align}
and
$T_3'(x)=2/(L_2-O_2)$ on $[(L_2+O_2)/2,L_2]$ and $0$ otherwise.
After substituting (\ref{eq:fem-approx}) into (\ref{eq:mean-field-1d}),
multiply both sides by $T_1(x)$ and $T_2(x)$ and then integrate. This yields
the following equations
\begin{displaymath}
  \int_0^{L_2} \rho'(x)D(x)T_1'(x)=0
\;\textrm{and}\;
  \int_0^{L_2} \rho'(x)D(x)T_2'(x)=0.
\end{displaymath}
From here it follows that
\begin{displaymath}
\int_0^{\frac{L_2-O_2}{2}}
\!\!\!\!
(\rho_\rr{u}T_0'+a T_1')T_1'
+
\int_{\frac{L_2-O_2}{2}}^{\frac{L_2+O_2}{2}}
(aT_1'+bT_2') \alpha T_1'=0
\end{displaymath}
and
\begin{displaymath}
\int_{\frac{L_2-O_2}{2}}^{\frac{L_2+O_2}{2}}
\!\!\!\!
(aT_1'+bT_2')\alpha T_2'
+\int_{\frac{L_2+O_2}{2}}^{L_2}(bT_2'+\rho_\rr{d}T_3') T_2=0.
\end{displaymath}
After integration, we obtain the next linear system
\begin{align}
  &\frac{1}{L_2-O_2}(a-\rho_\rr{u})+\frac{\alpha}{O_2}(b-a)=0,\nonumber\\
  &\frac{\alpha}{O_2}(b-a)-\frac{1}{L_2-O_2}(\rho_\rr{d}-b)=0.\nonumber
\end{align}
yielding
\begin{equation}
  \label{eq:fem-coeff-3}
 a=\frac{\rho_\rr{u}+\rho_\rr{d}+\rho_\rr{u}\beta}{2+\beta}
\;\;\textrm{ and }\;\;
 b=\frac{\rho_\rr{u}+\rho_\rr{d}+\rho_\rr{d}\beta}{2+\beta},
\end{equation}
with
\begin{equation}
\label{eq:fem-coeff-1}
\beta=\frac{O_2}{\alpha(L_2-O_2)}.
\end{equation}

\begin{figure}[t]
  \centering
  \includegraphics[width=0.45\textwidth]{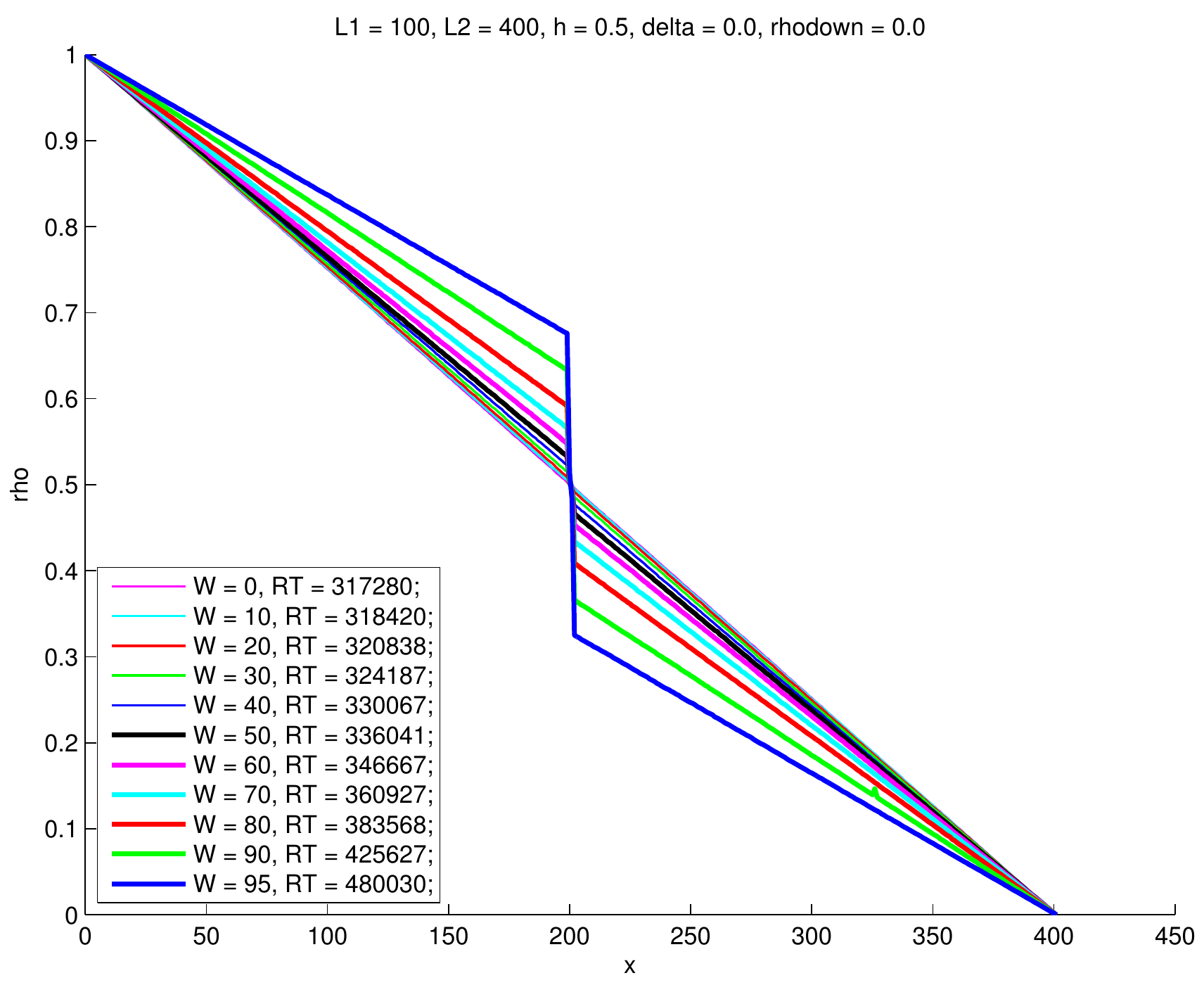}
  \includegraphics[width=0.45\textwidth]{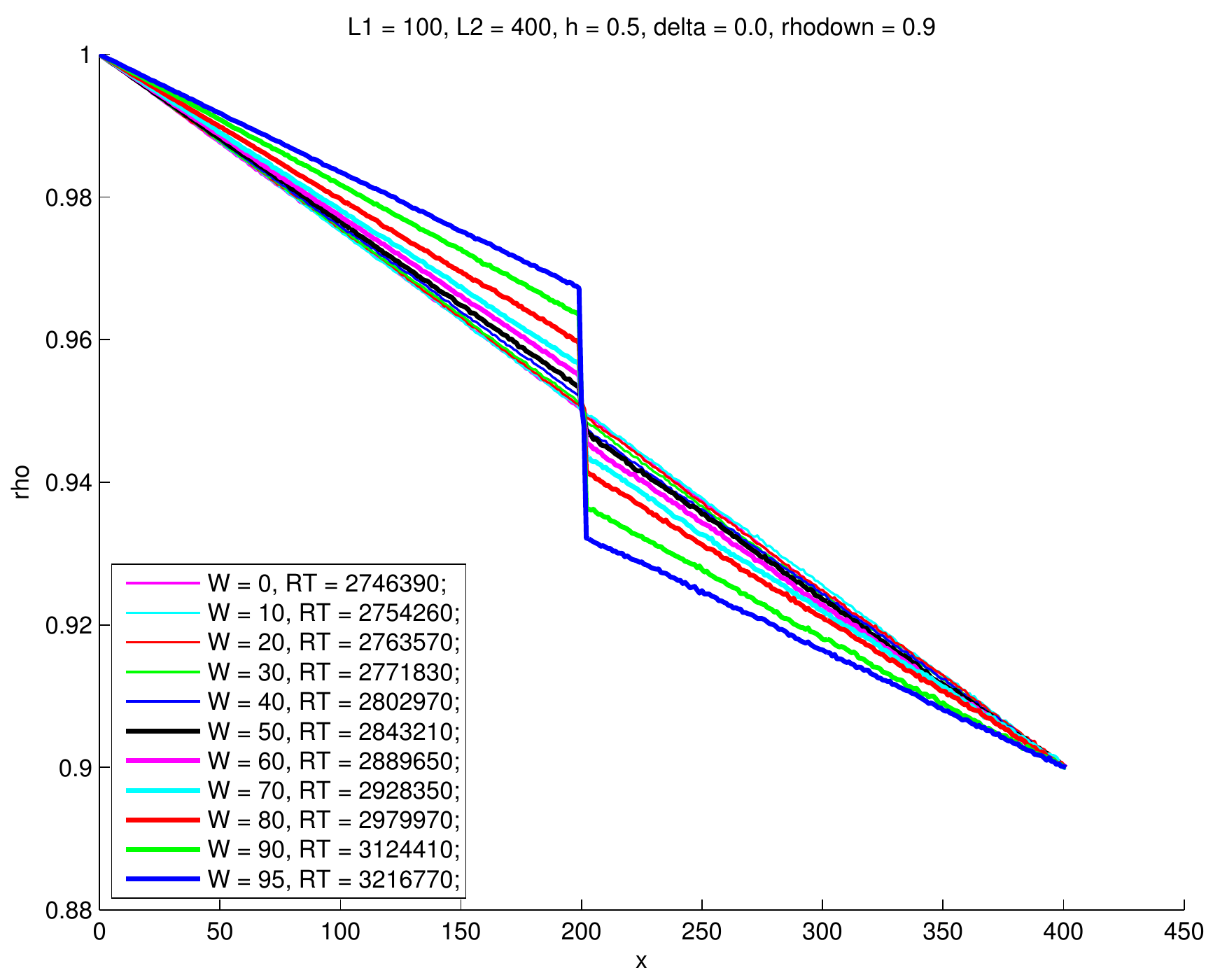}
  \caption{Density profile obtained by averaging the
  2D lattice simulation: comparison for different $W$.
    Parameters: $L_1=100$, $L_2=400$, $h=0.5$, $\delta=0$,
    $\rho_\rr{u}=1$,
    $\rho_\rr{d}=0$ (top) and $\rho_\rr{d}=0.9$ (bottom),
    $O_2=3$,
    and $W$ as listed in the inset.
    In the inset we have also listed the residence time
    data discussed in Section~\ref{s:onert}.
}
  \label{fig:width-profile-comparison-drift-00}
\end{figure}

We remark that
the deviations in the density profile from the straight line are
symmetric.
See e.g. Figure~\ref{fig:width-profile-comparison-drift-00} for an
example simulation.
Indeed,
by summing the coefficients in (\ref{eq:fem-coeff-3}) we obtain
$a+b=\rho_\rr{u}+\rho_\rr{d}$ and, hence.
\begin{displaymath}
    a-\frac{\rho_\rr{u}+\rho_\rr{d}}{2}
=\frac{(\rho_\rr{u}-\rho_\rr{d})\beta}{4+2\beta}
\end{displaymath}
and
\begin{displaymath}
\frac{\rho_\rr{u}+\rho_\rr{d}}{2}-b
=\frac{(\rho_\rr{u}-\rho_\rr{d})\beta}{4+2\beta}.
\end{displaymath}

Having obtained the analytical expression for $\rho(x)$, we can compute
the 1D Mean Field residence time approximation
by using \eqref{eq:residence-time}.
Indeed, some simple algebra yields
\begin{equation}
  \label{eq:drift-zero-RT-estimate}
  R=\frac{(\rho_\rr{u}+\rho_\rr{d})L_2(L_2-O_2)(2+\frac{2O_2}{\alpha(L_2-O_2)})}{\rho_\rr{u}-\rho_\rr{d}}
\end{equation}
where, we recall, $\alpha=F(h)(L_1-W)/L_1$.
In the case $\alpha=1$, i.e.,
no obstacle, the expression of the residence time simplifies to
\begin{equation}
  \label{eq:drift-zero-RT-estimate-no-obstacle}
  R=\frac{(\rho_\rr{u}+\rho_\rr{d})2L_2^2}{\rho_\rr{u}-\rho_\rr{d}},
\end{equation}
which is an agreement with \cite[equation~(5.39)]{Rutger}.

\begin{figure}[t]
  \centering
  \includegraphics[width=0.45\textwidth]{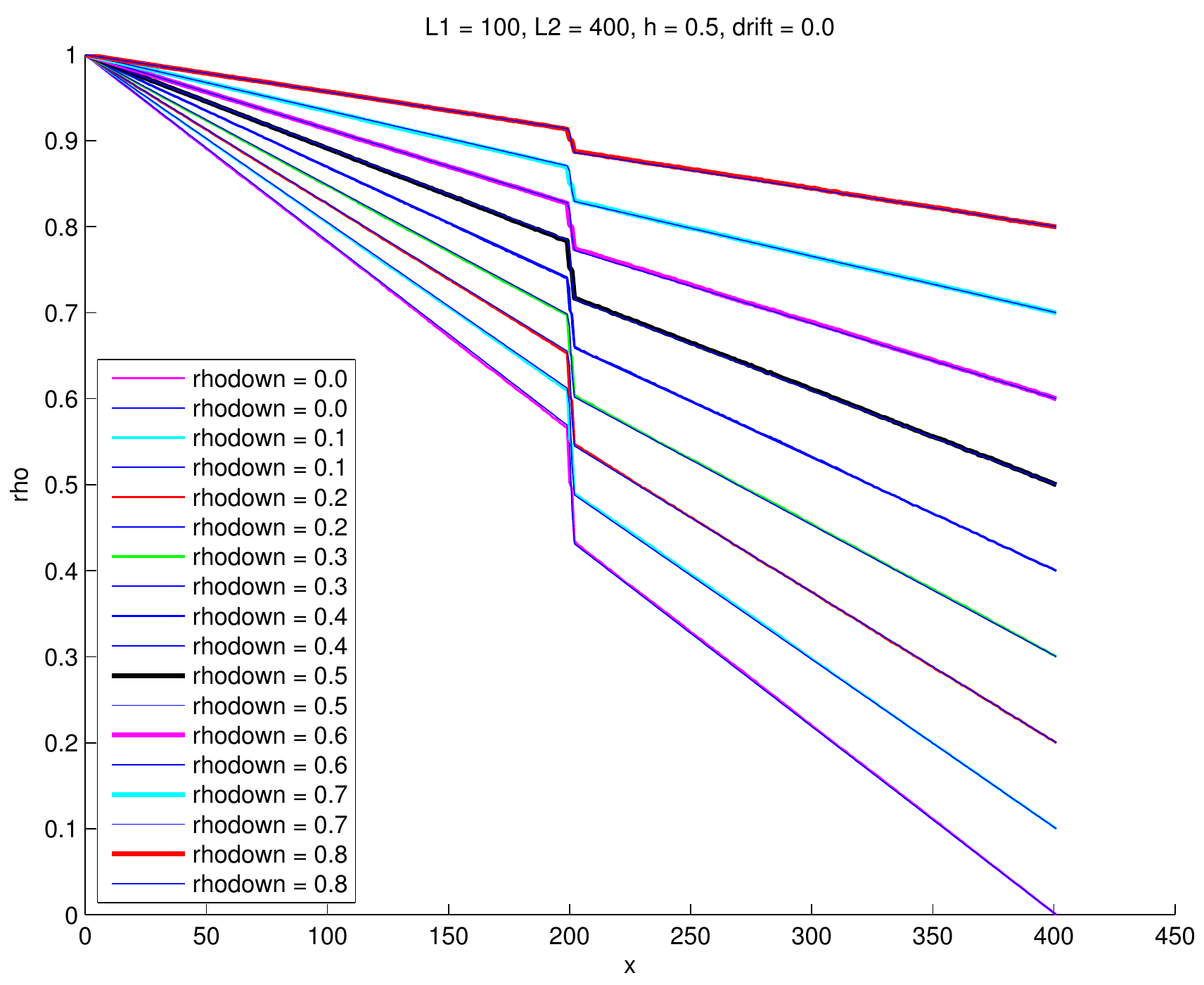}
  \caption{Comparison between the density profile obtained by
    averaging the 2D lattice simulation and the analytical solution of
    the 1D mean field equation.
    Parameters: $L_1=100$, $L_2=400$, $h=0.5$, $\delta=0$,
    $W=70$,
    $O_2=3$, $\rho_\rr{u}=1$, and $\rho_\rr{d}$ as listed in the
    inset.
    For the 1D model the tortuosity coefficient
    has been chosen equal to $0.45$ for all the values of $\rho_\rr{d}$.
    Thick lines correspond to Monte Carlo data for the lattice model
    and thin lines correspond to the analytical solution of the
    1D model.}
  \label{fig:compare1d}
\end{figure}

We note the following:
according to (\ref{eq:drift-zero-RT-estimate}), the residence time
  increases with increasing value of $\rho_\rr{d}$.
  Additionally, the effect of $W$ on the residence time disappears when $L_2$ goes to infinity.
Moreover,
from (\ref{eq:drift-zero-RT-estimate}), the residence time uniformly
  increases as $W$ increases.  Note also that the $W$ dependence can
  be also seen purely as $W/L_1$. This is a limitation of our simple
  approximation to the diffusion coefficient, since in our simulations
  we see an effect of different values of $W$ on the residence time,
  even with the
  same $W/L_1$ ratio (see Section~\ref{s:onert}).

\subsection{Density profile}
\label{s:onedp}
Now, we discuss how the density profile obtained from
\eqref{eq:fem-approx} compares to the one obtained by averaging the 2D
Monte Carlo simulation. The results are shown in
Figure~\ref{fig:compare1d} in the case $W=70$.
The parameters we used in the computation are listed in the caption.

The match between the Monte Carlo and the analytical result
is perfect. For the 1D model we had to
optimize on the tortuosity coefficient by choosing
$F=0.45$ for this comparison, but we stress that the same value
has been used for all the choices of the bottom boundary
density plotted in the picture.
Although this value resulted in a good match,
the question of the explicit dependence $F(h)$ still remains open.

The size of the jump in the averaged density profile, which can be
observed in the figure, obviously
depends on the width of the obstacle. This dependence is
analyzed in Figure~\ref{fig:width-profile-comparison-drift-00},
where we plot the averaged Monte
Carlo density
profile for the 2D lattice model for different values of $W$.
The two plots show our results for $\rho_\rr{d}=0$ (top panel)
and $\rho_\rr{d}=0.9$ (bottom panel). It is worth remarking that, as
we expected, in both cases the size of the jump increases with
the obstacle width. But we stress that the qualitative behavior of
the graph does not change with $\rho_\rr{d}$. This fact is particularly
relevant and it is key in our explanation for the different behaviors that we
shall find in the biased (not zero drift) case.

\begin{figure}[t]
  \centering
  \includegraphics[width=0.45\textwidth]{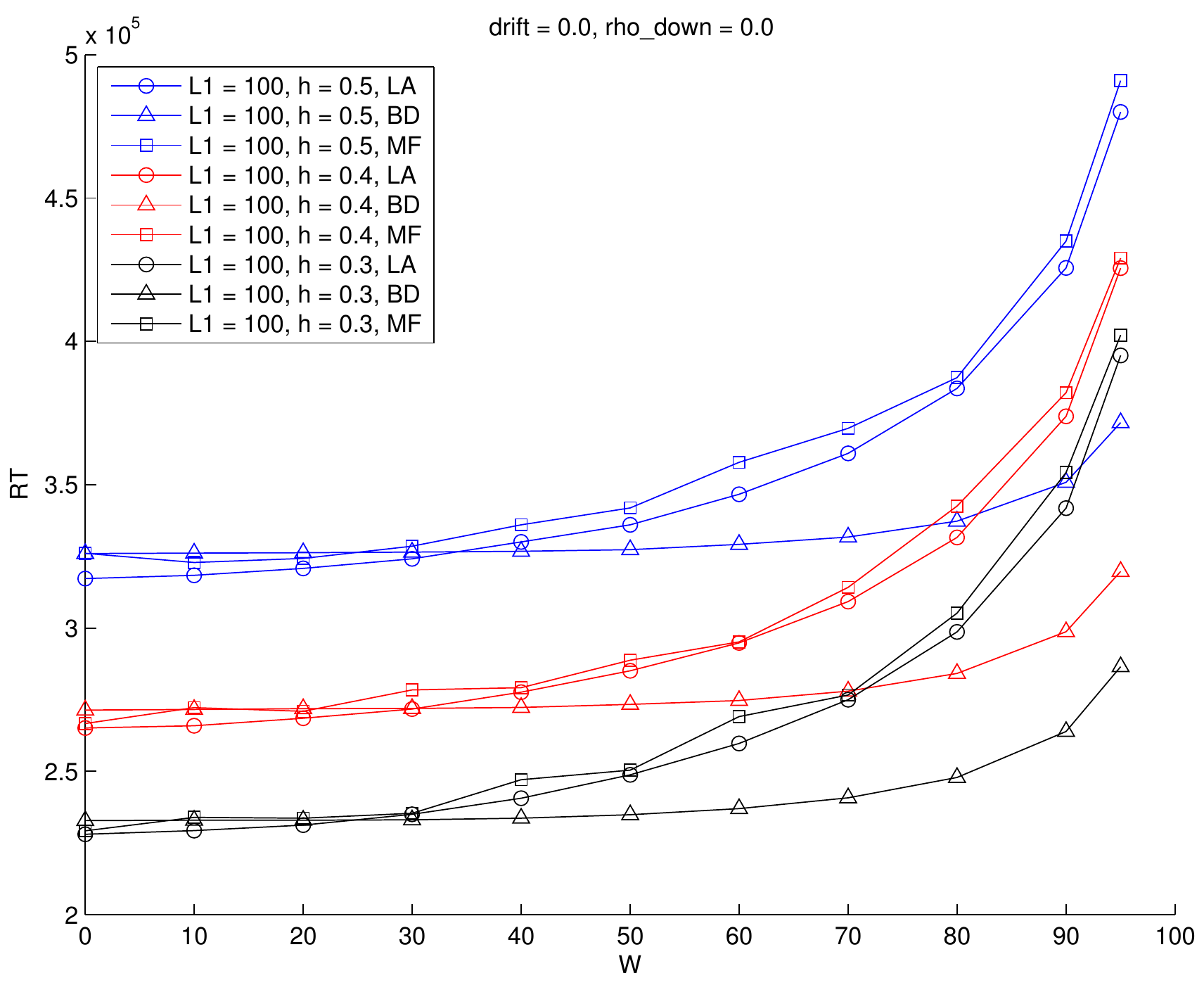}
 \caption{The BD and MF approximations to the actual measured mean residence
 time (labeled LA).
 Parameters: $L_1=100$, $L_2=400$, $\delta=0$,
 $\rho_\rr{u}=1$, $\rho_\rr{d}=0$,
 $O_2=3$,
 and $h$ as listed in the inset.}
  \label{fig:bd-mf-0.0}
\end{figure}

\begin{figure}[t]
  \centering
  \includegraphics[width=0.45\textwidth]{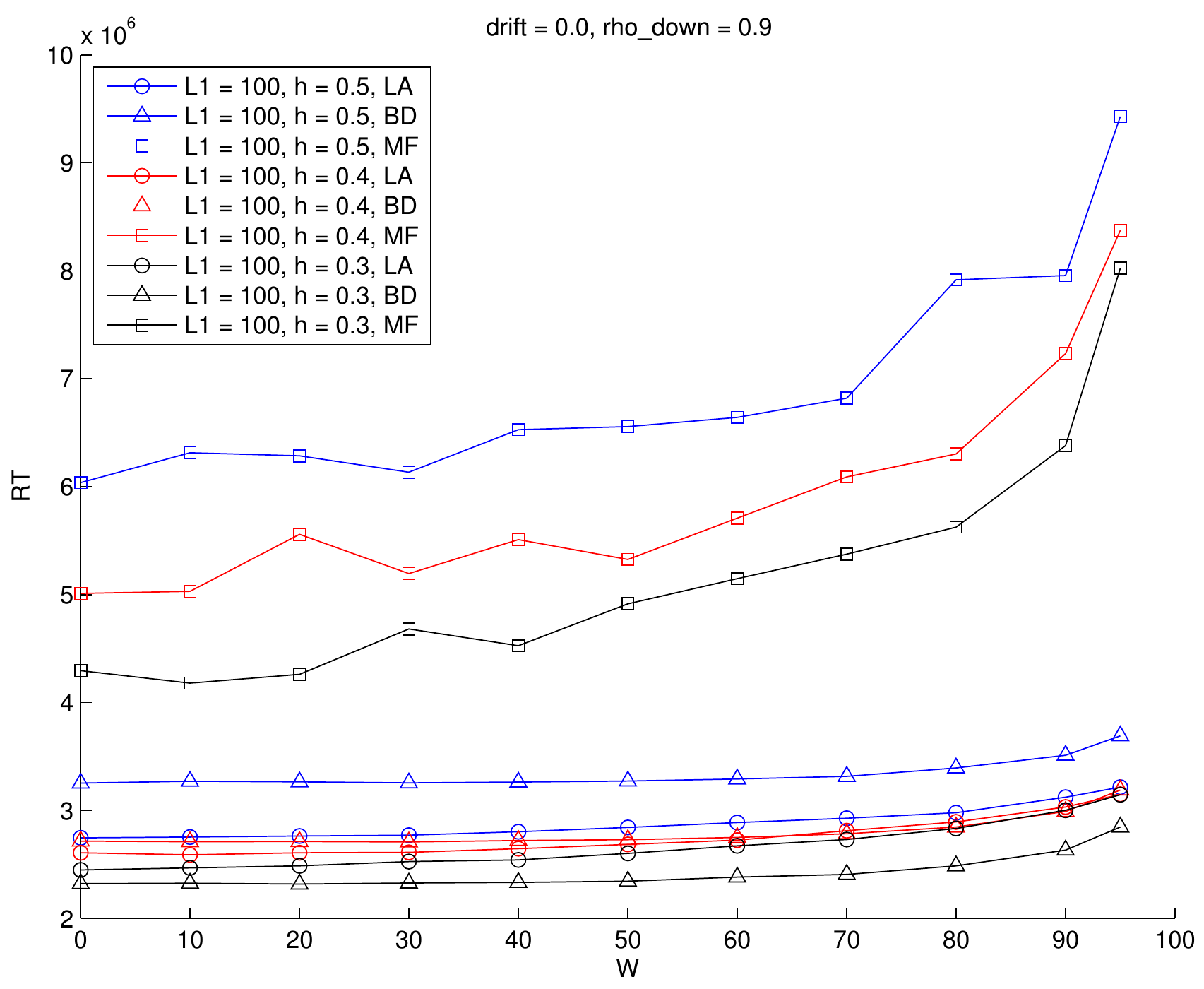}
  \caption{The BD and MF approximations to the actual measured mean residence
  time (labeled LA).
 Parameters: $L_1=100$, $L_2=400$, $\delta=0$,
 $\rho_\rr{u}=1$, $\rho_\rr{d}=0.9$,
 $O_2=3$,
 and $h$ as listed in the inset.}
  \label{fig:bd-mf-0.9}
\end{figure}

\subsection{Residence time}
\label{s:onert}
The above discussion shows that the 2D stationary density profile
can be found by averaging the Monte Carlo data for the 2D lattice
model or by solving the Mean Field model \eqref{MF}. Moreover,
by averaging along the horizontal axis this 2D profile, we find
a 1D profile which can be perfectly fitted with the 1D
model proposed in Section~\ref{onedim} by choosing the correct
tortuosity coefficient. Such a 1D density profile can be
used as an input to estimate the residence time.

This estimate can be achieved via the Mean Field
approximation provided in \eqref{rt-mf}. In \cite{Rutger}
a different approach, base on a Birth--and--Death model
has been also proposed and thoroughly discussed in absence
of obstacles. The main idea is that of predicting the residence
time via a one--dimensional model in which a particle perform
a simple random walk in the vertical direction with jumping
probabilities defined in terms of the stationary density
profile measured for the 2D lattice model.
In particular it has been
deduced the prediction \cite[equation~(4.20)]{Rutger}
for the residence time
based on the Birth--and--Death model defined in
\cite[equation~(5.28)]{Rutger}.
In that paper, due to the
absence of obstacle, the reduction to 1D is rather obvious, since
the density profile does not depend on the horizontal coordinate.
As already remarked above,
in the present case we shall use this theory starting from the
horizontally averaged density $\tilde\rho$ as in equation~\ref{rt-mf}.

In Figure~\ref{fig:bd-mf-0.0} and Figure \ref{fig:bd-mf-0.9}, we
compare the Monte Carlo measurement of the residence time to the
Birth--and--Death and Mean Field estimates based on the
horizontally averaged 1D
density profile of a 2D simulation of a flow through a strip with an
obstacle in the middle.  On the horizontal axis we have the obstacle
width and on the vertical axis the mean residence time. The formulas
for the both residence time estimates can be found in \cite{Rutger},
more precisely, see
\cite[equation~(5.32)]{Rutger}
and
\cite[equation~(5.39)]{Rutger}
respectively
for the BD and the MF approximation.

As we can see, the quality of the approximation is influenced heavily
by the value of $\rho_\rr{d}$. For $\rho_\rr{d}=0$, the MF approximation
works very well, while the BD approximation gets worse when
the width of the obstacle is increased.
For
$\rho_\rr{d}=0.9$, the MF approximation overestimates by a lot, while the
BD approximation is a bit better, but still not very precise.
This result is consistent with what it has been found in \cite{Rutger}
in absence of obstacle: in absence of drift, provided $h$ is large
enough (here we are using $h=0.5$), the BD prediction is better
than the MF one in those situations in which clogging is present. There,
in absence of obstacles, clogging was introduced by increasing the
value of the bottom boundary density.

From this it follows that we can't expect to get great residence
time  estimates
based on the analytical solution of our 1D model for the case of zero
drift. But we can still hope to reproduce the density profiles well.

As a final remark, on which we shall come back in
the discussion Section~\ref{Discussion} in connection with the
results we will find in the not zero drift situation, we note that
the behavior of the residence time with the obstacle width is absolutely
trivial. Indeed, it stays more or less constant till half the horizontal
width is reached, then it increases sharply.

\section{Non--zero Drift Case}
\label{Results-nz}
We consider the lattice model introduced in Section~\ref{lattice}
on the lattice strip of size $L_1\times L_2$ in presence of drift,
namely, for $\delta>0$.
Our simulations will be run mainly for the same parameters
as those used in Section~\ref{Results}. Details will be given in
the figure captions.

In this case, since particles do experience an
external drift, we expect that the stationary density
profile will depend on the horizontal lattice coordinate.
For this reasons our discussion will rely exclusively on
the Monte Carlo simulation of the 2D lattice model
introduced in Section~\ref{lattice}.

\subsection{Density profile}
\label{dp-dr}
The density profile is measured for the 2D lattice model,
see also the comments Section~\ref{lattice},
by averaging the occupation number at stationarity.
Our results are plotted in
Figure~\ref{fig:2D-density-drift-nonzero}, where we used
the parameters
$L_1=100$, $L_2=400$, $h=0.5$, $\delta=0.05$, $\rho_\rr{u}=1$,
$\rho_\rr{d}=0,0.9$, $O_2=3$, and $W=85,90,95$; recall
the obstacle is placed in the middle of the strip.
The main features are: the presence of a jump across
the obstacle and the dependence of the profile on the
horizontal coordinate.

A deeper insight in the structure of the density profile can
be reached by looking at the horizontally averaged densities.

\begin{figure*}[t]
  \centering
  \begin{tabular}{ll}
    \includegraphics[width=0.3\textwidth]{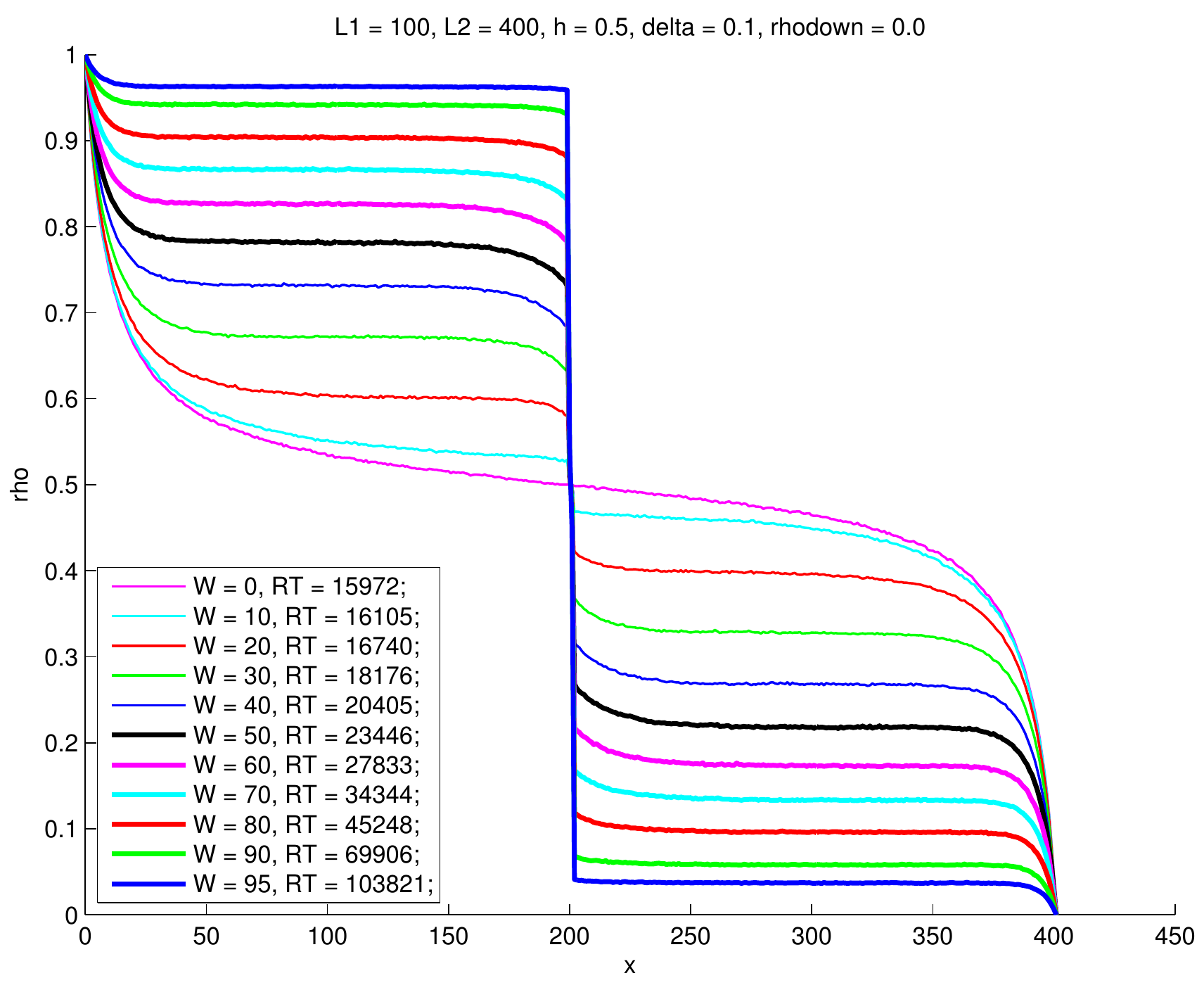}
    &
      \includegraphics[width=0.3\textwidth]{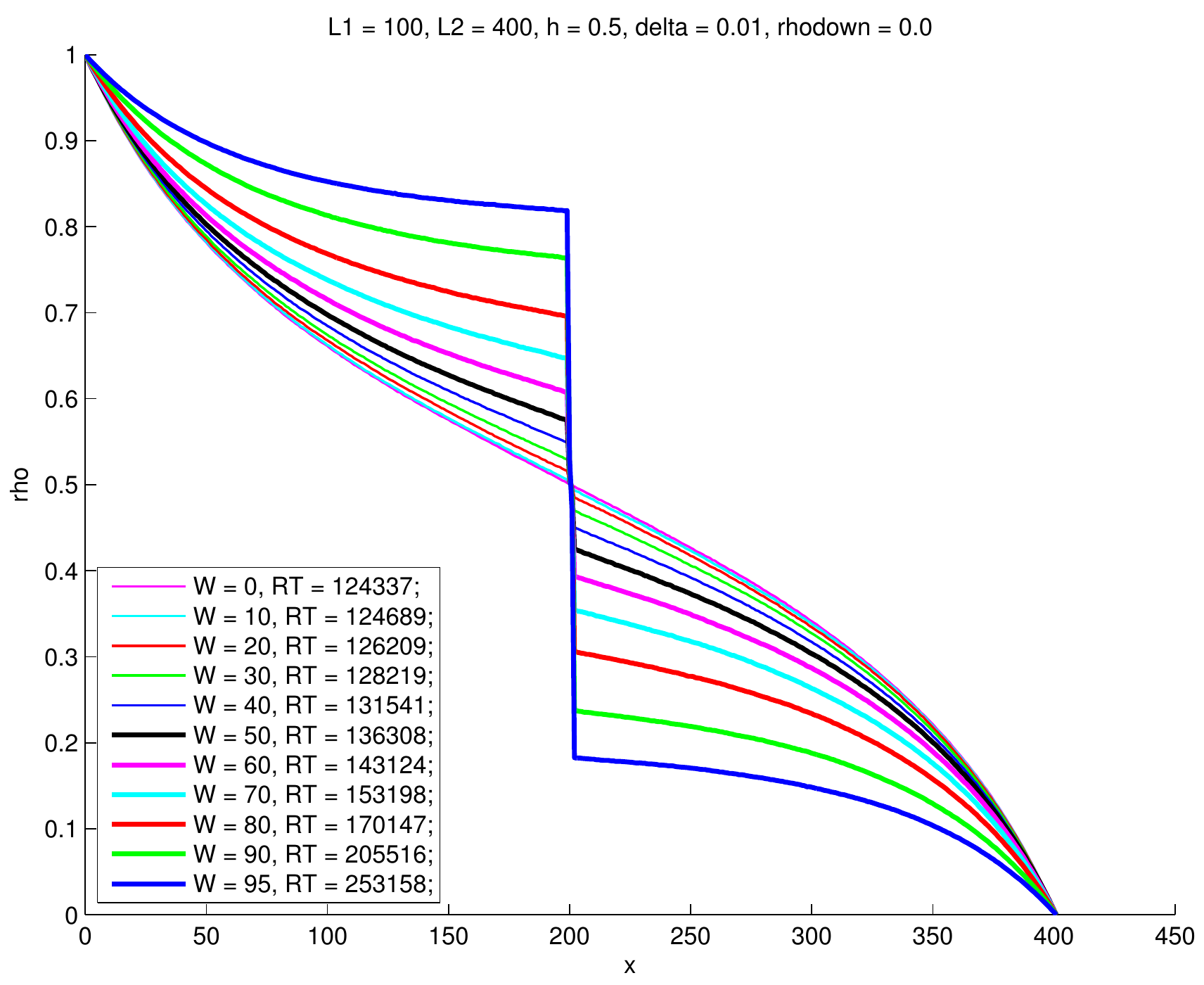}\\
    \includegraphics[width=0.3\textwidth]{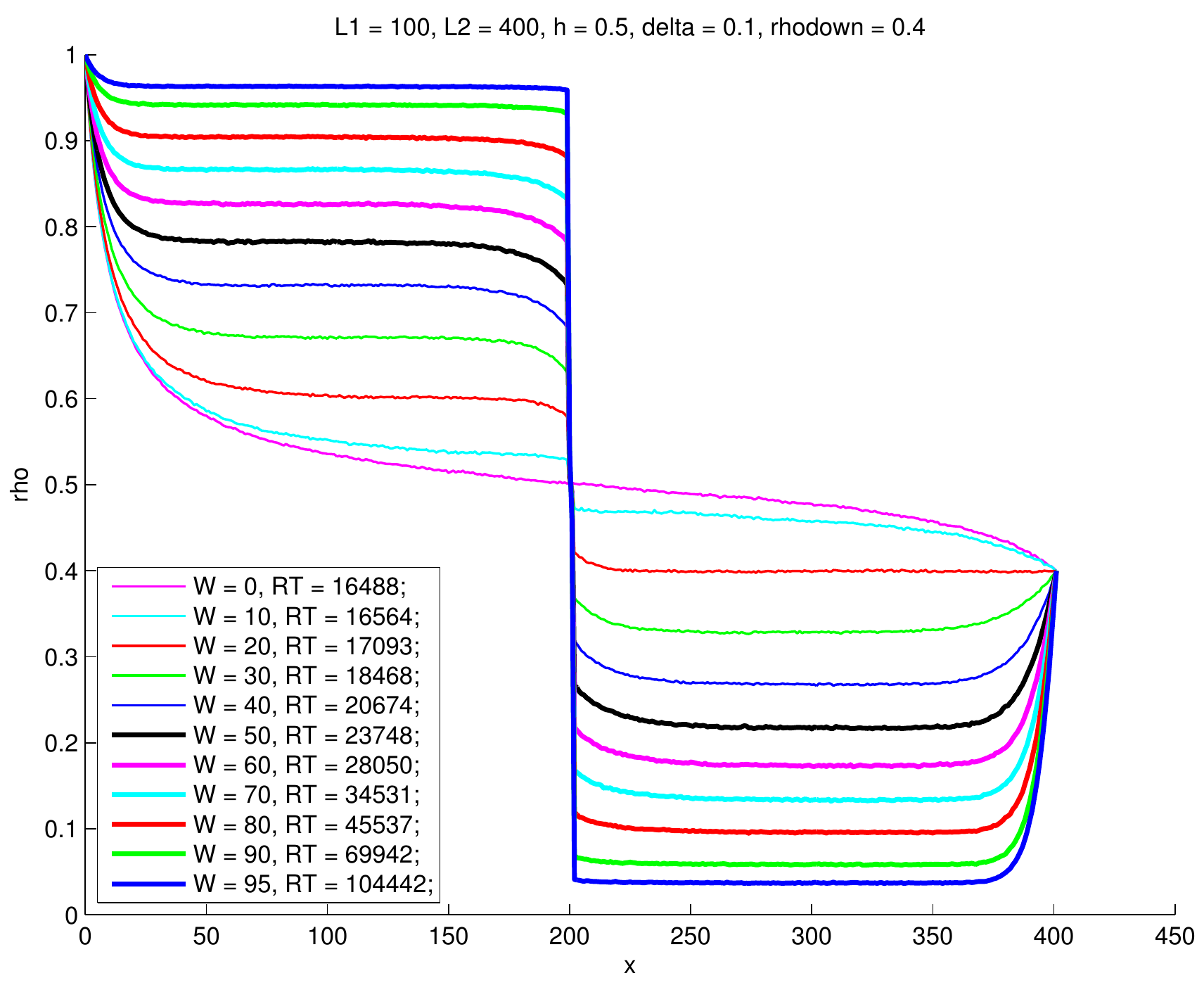}
    &
      \includegraphics[width=0.3\textwidth]{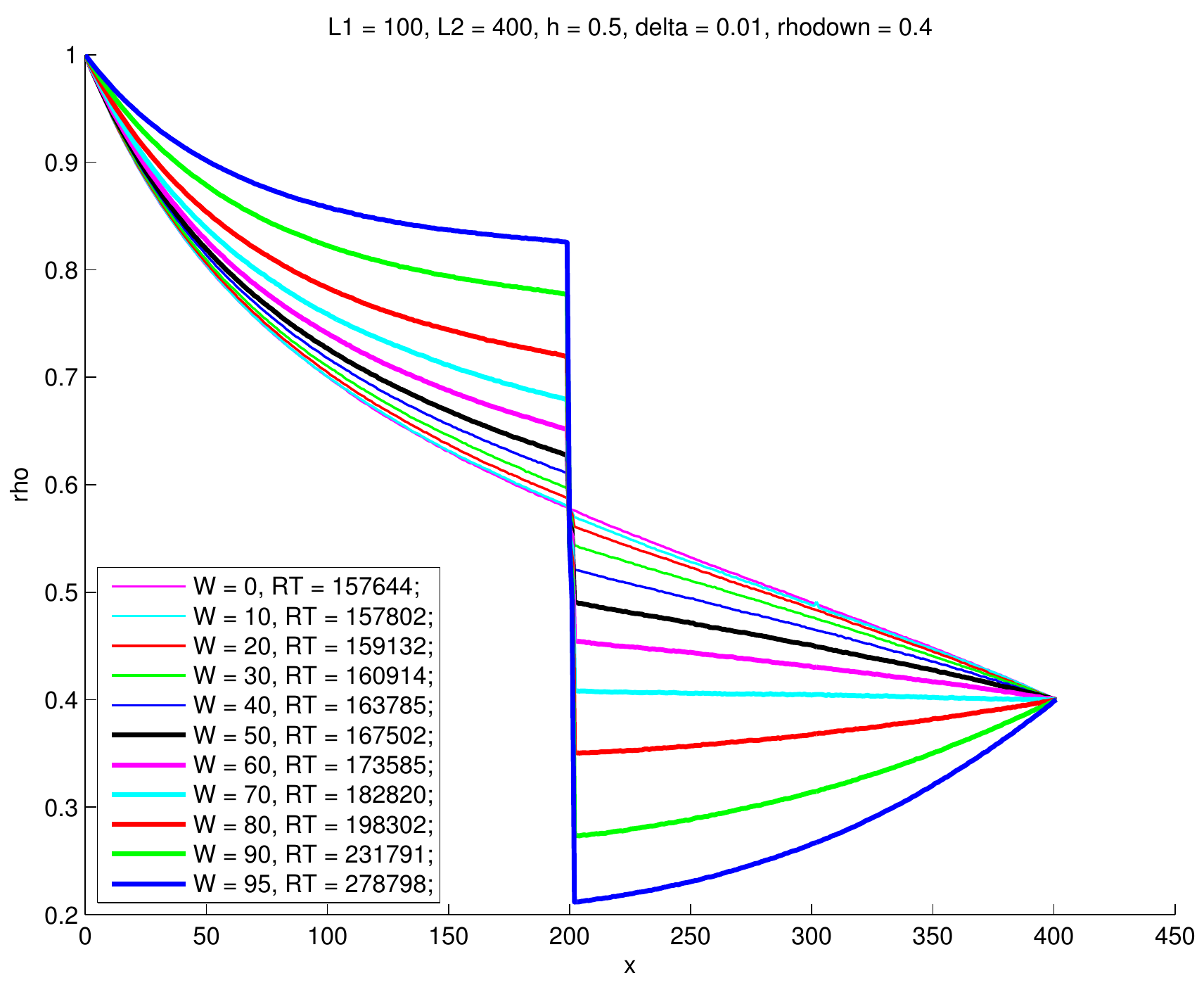}\\
    \includegraphics[width=0.3\textwidth]{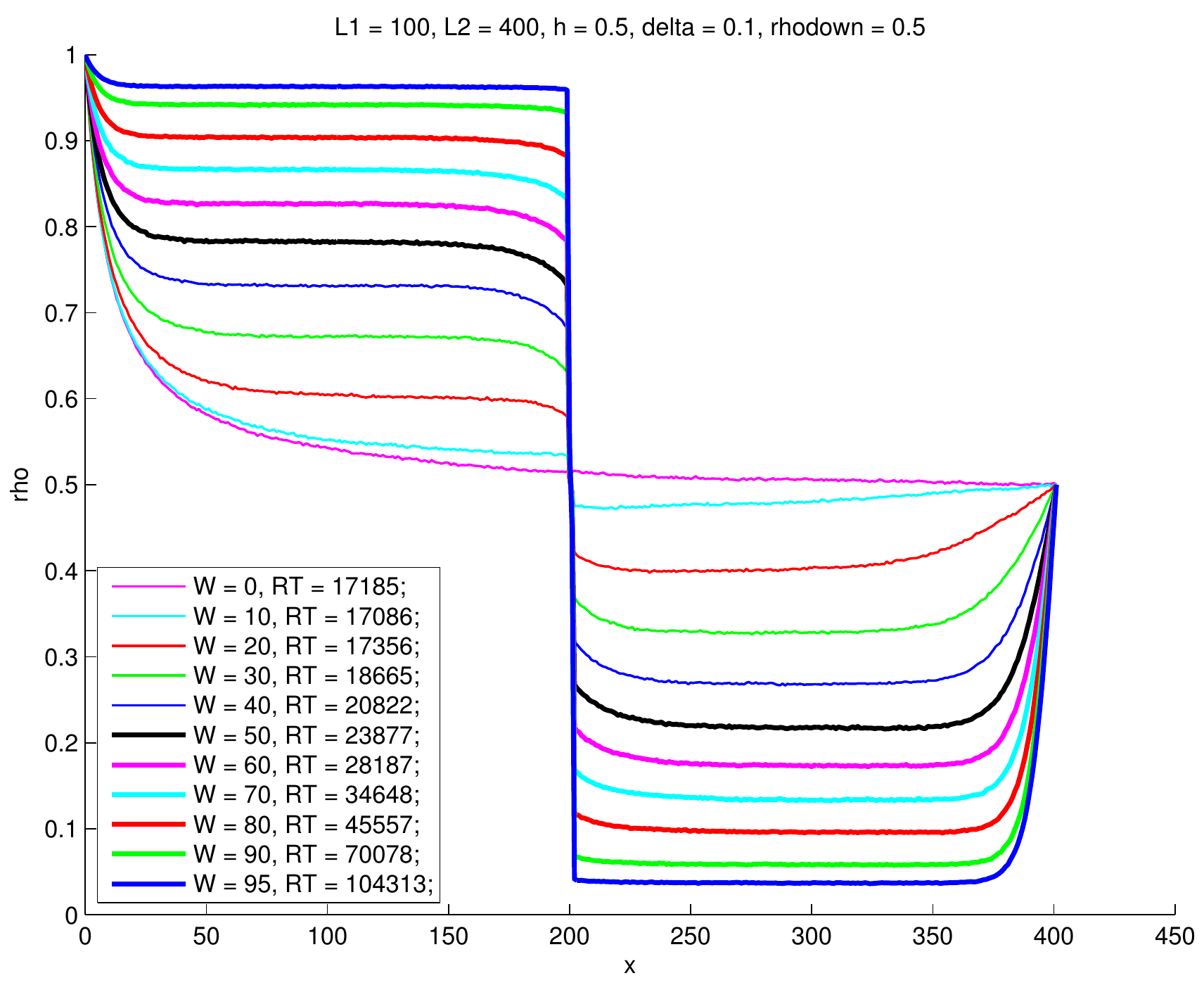}
    &
      \includegraphics[width=0.3\textwidth]{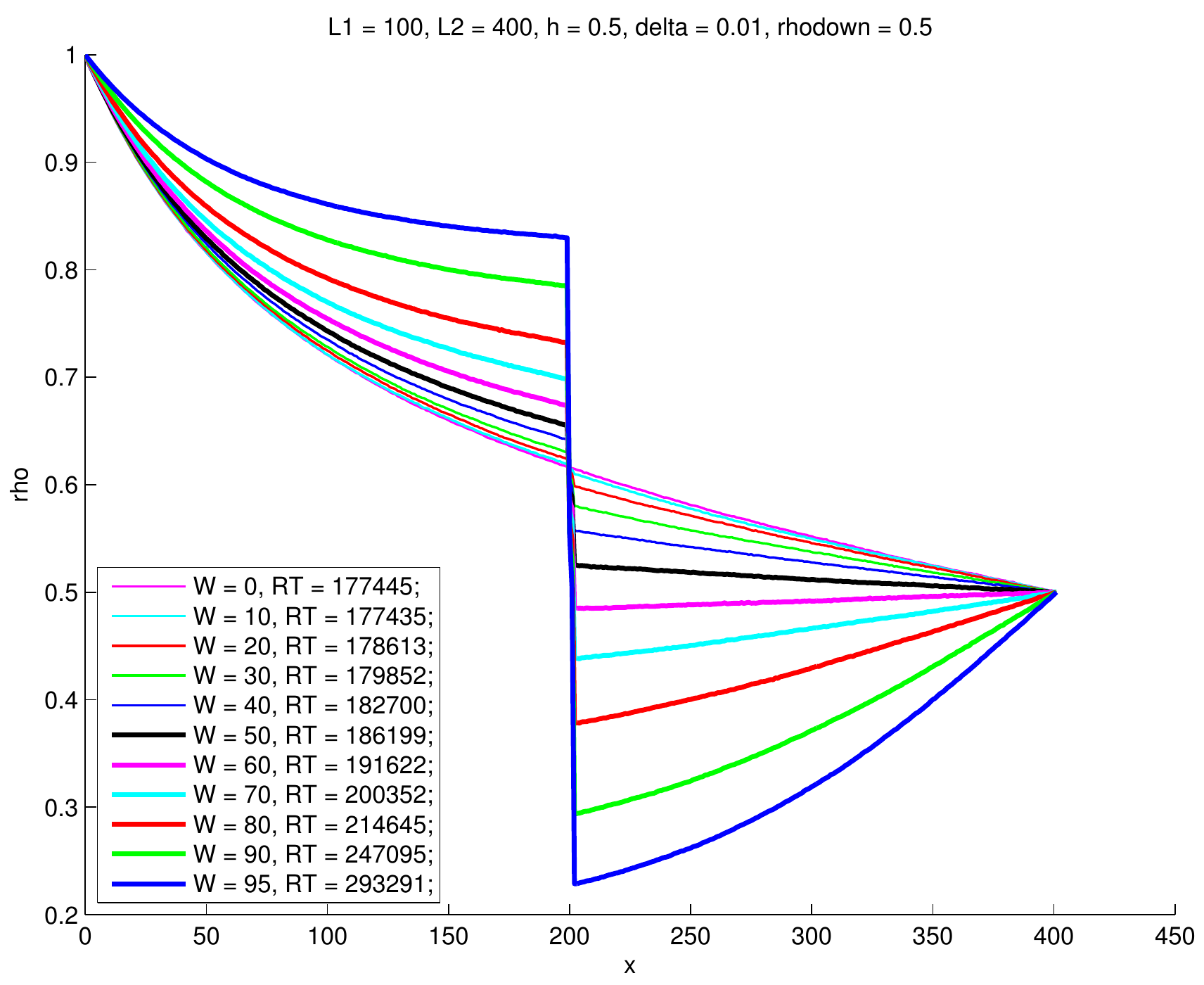}\\
    \includegraphics[width=0.3\textwidth]{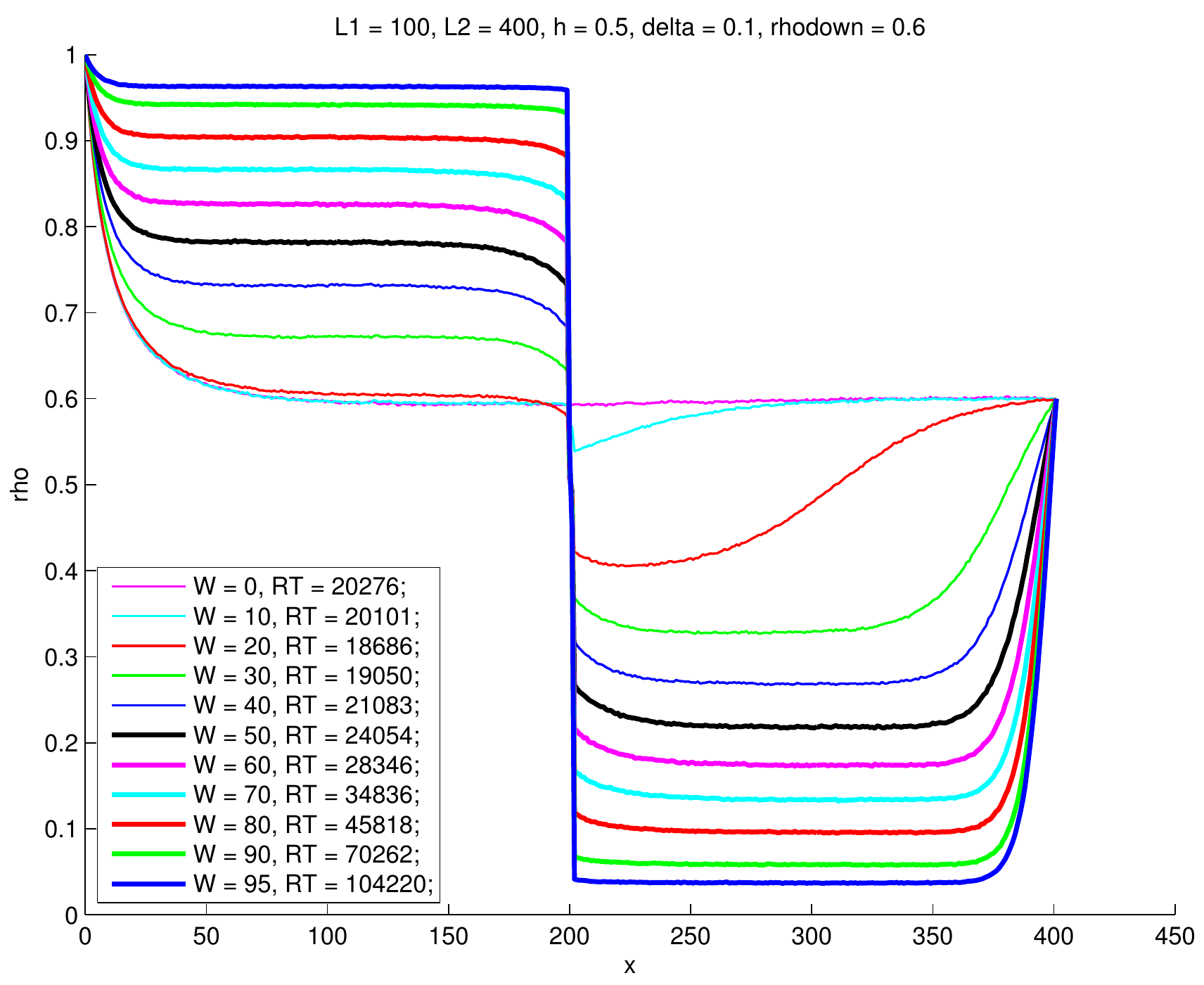}
    &
      \includegraphics[width=0.3\textwidth]{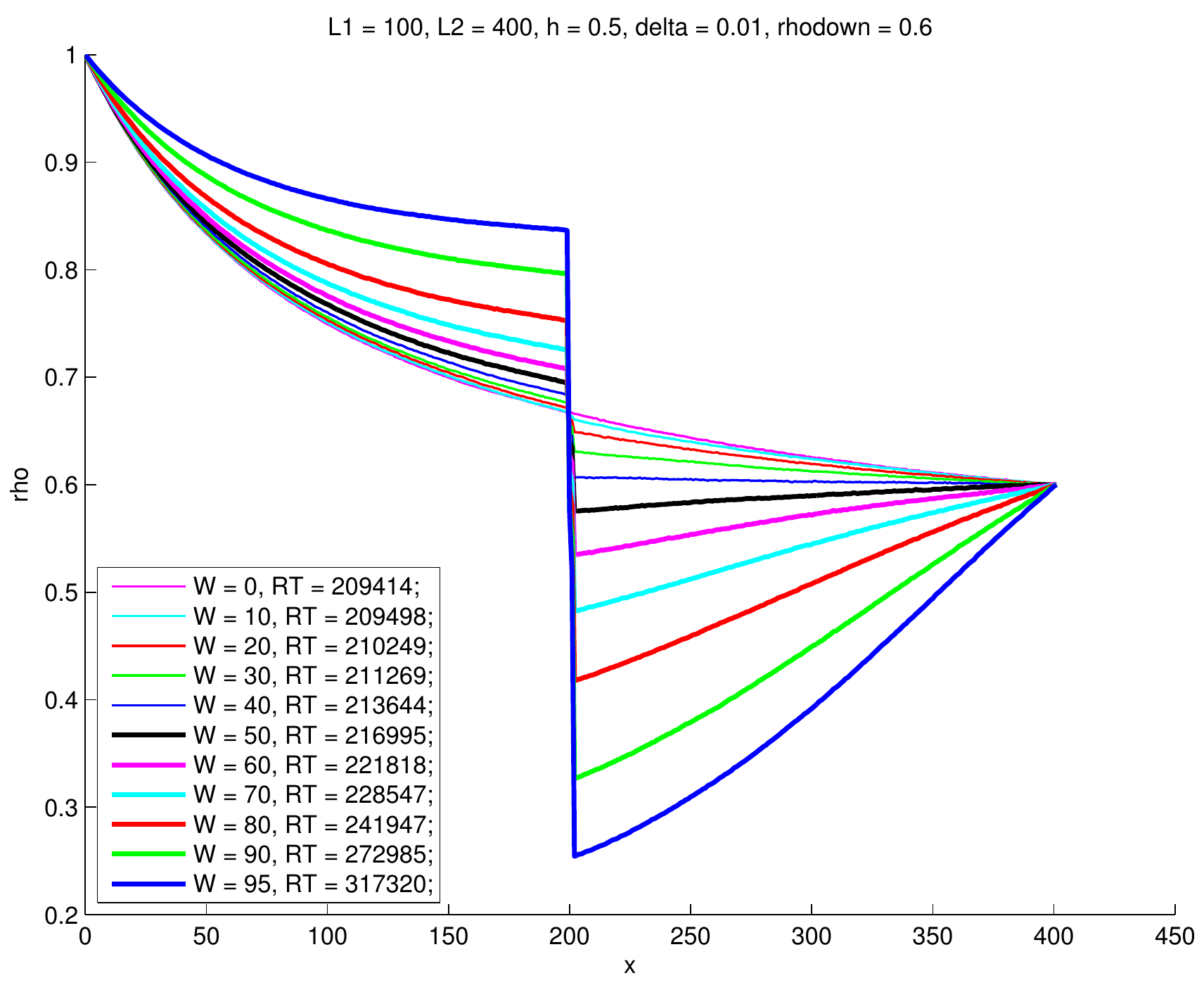}\\
  \end{tabular}
  \caption{Density profile obtained by averaging the
  2D lattice simulation: comparison for different $W$.
  The lattice size is $100\times 400$, $h=0.5$,
  $\drift=0.1$ (left) and $\drift=0.01$ (right),
  $\rho_\rr{u}=1$,
  $\rho_\rr{d}=0,0.4,0.5,0.6$ from the top to the bottom,
  and $O_2=3$.
  The residence time measured in the different cases has been reported in the
  inset.
}
  \label{fig:width-profile-comparison-drift-01-1}
\end{figure*}

\begin{figure*}[t]
  \centering
  \begin{tabular}{ll}
    \includegraphics[width=0.3\textwidth]{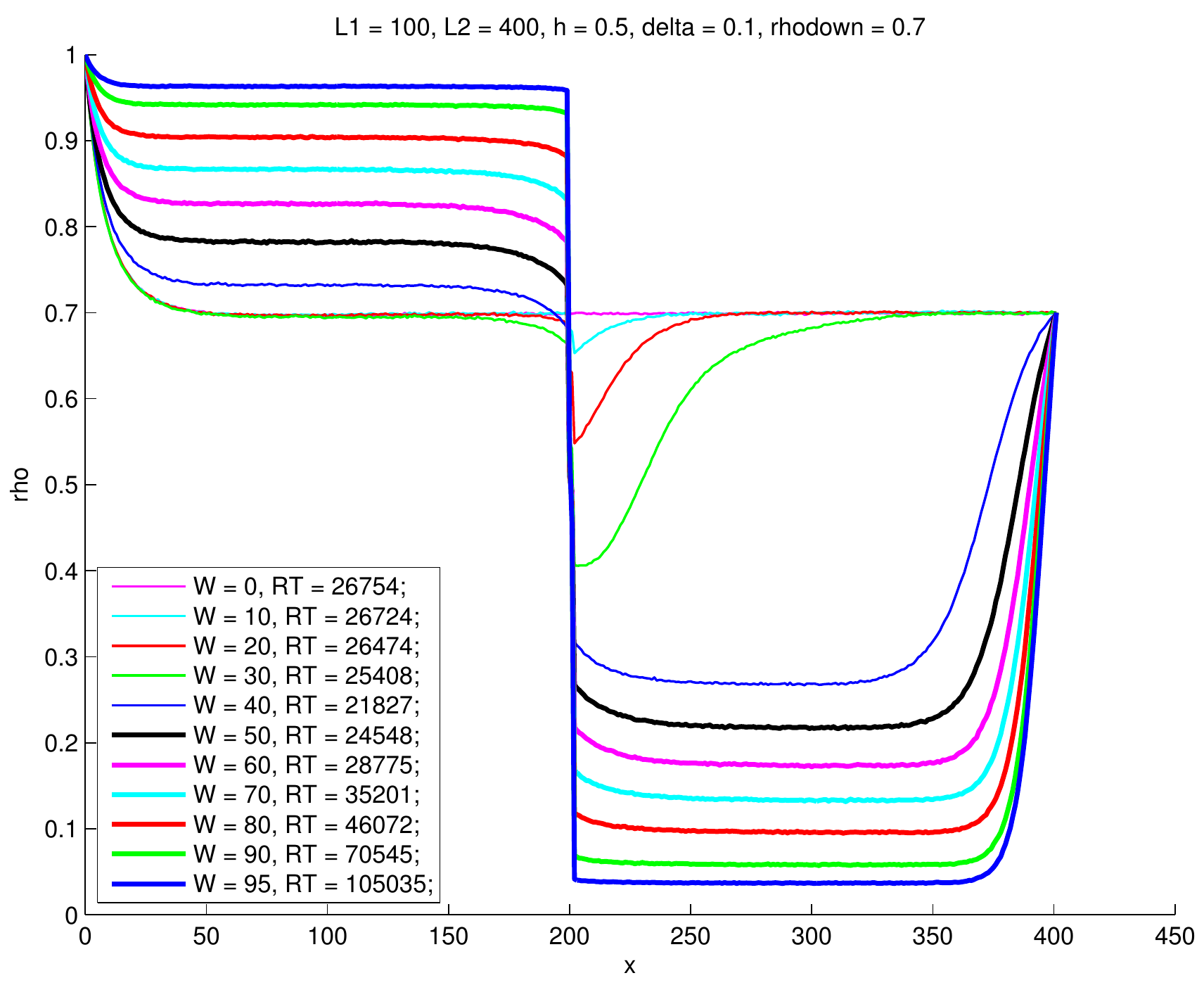}
    &
      \includegraphics[width=0.3\textwidth]{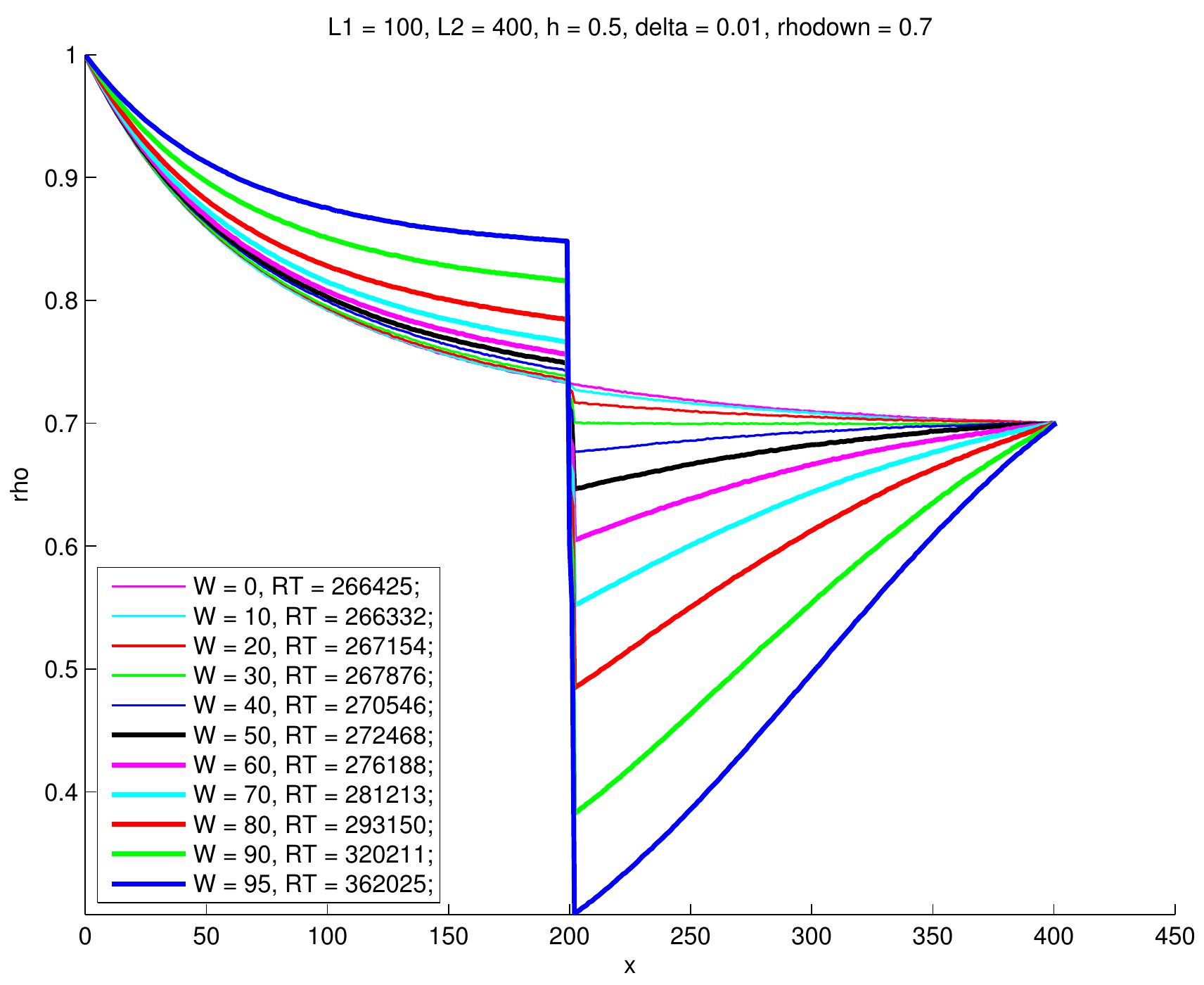}\\
    \includegraphics[width=0.3\textwidth]{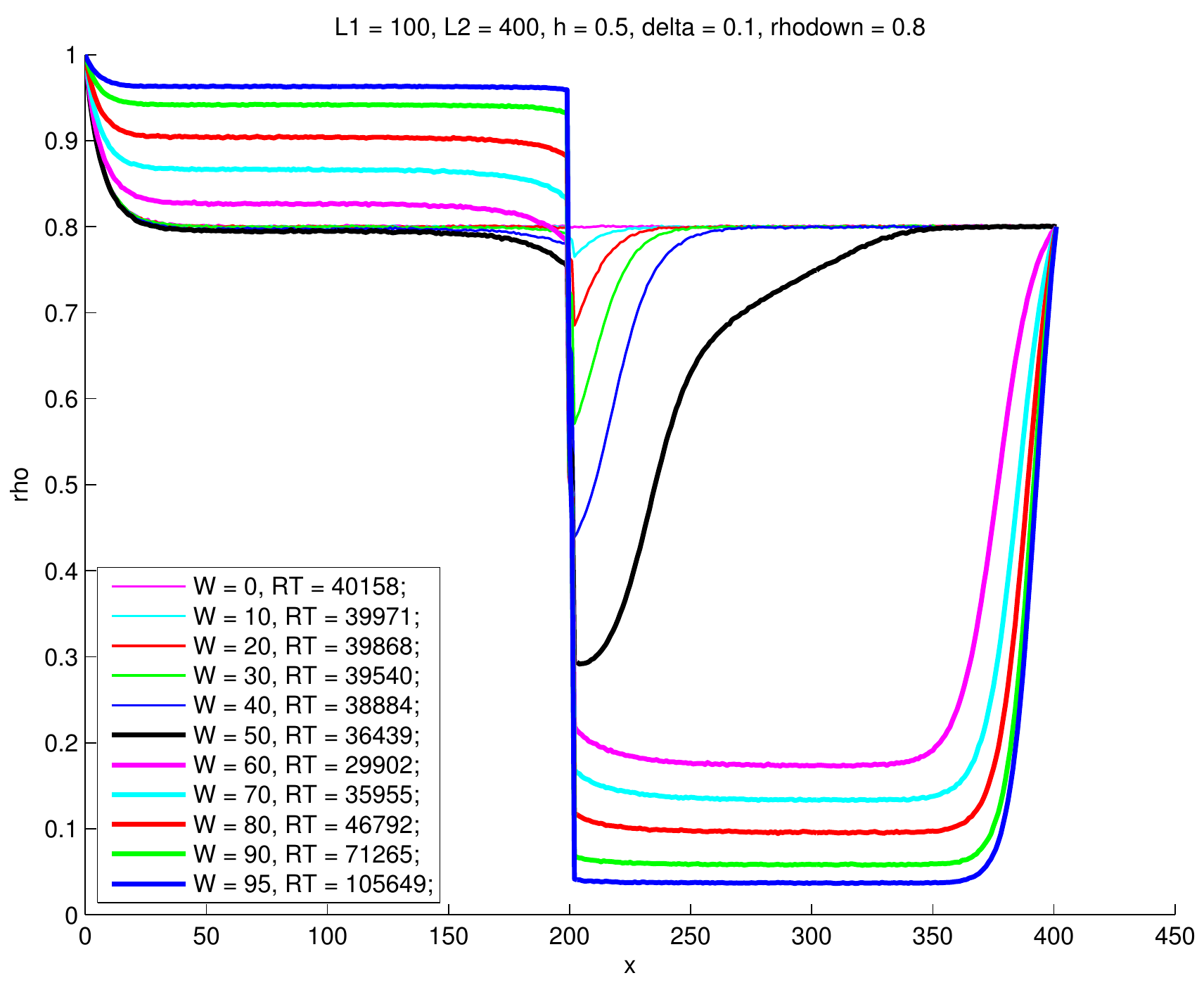}
    &
      \includegraphics[width=0.3\textwidth]{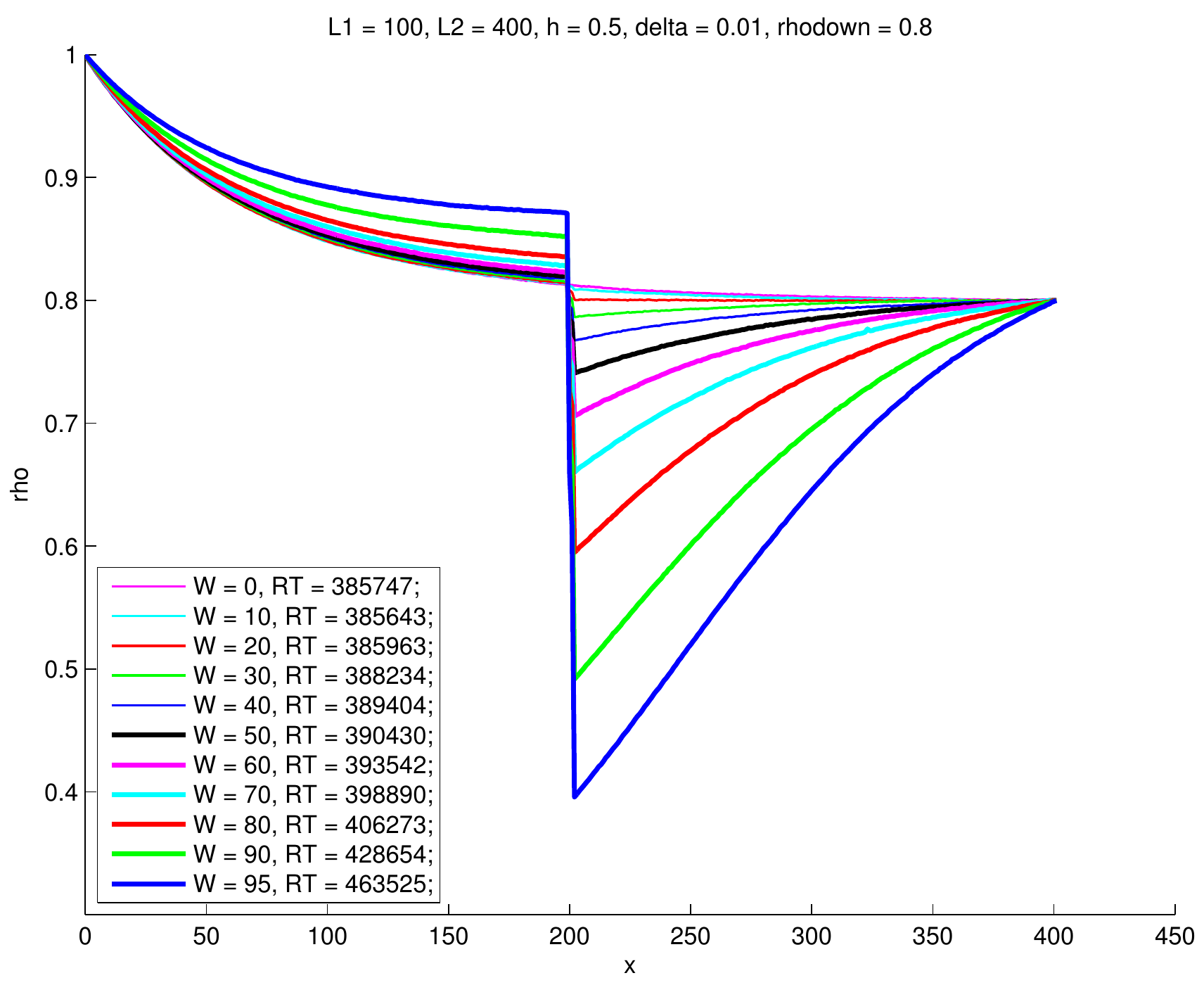}\\
    \includegraphics[width=0.3\textwidth]{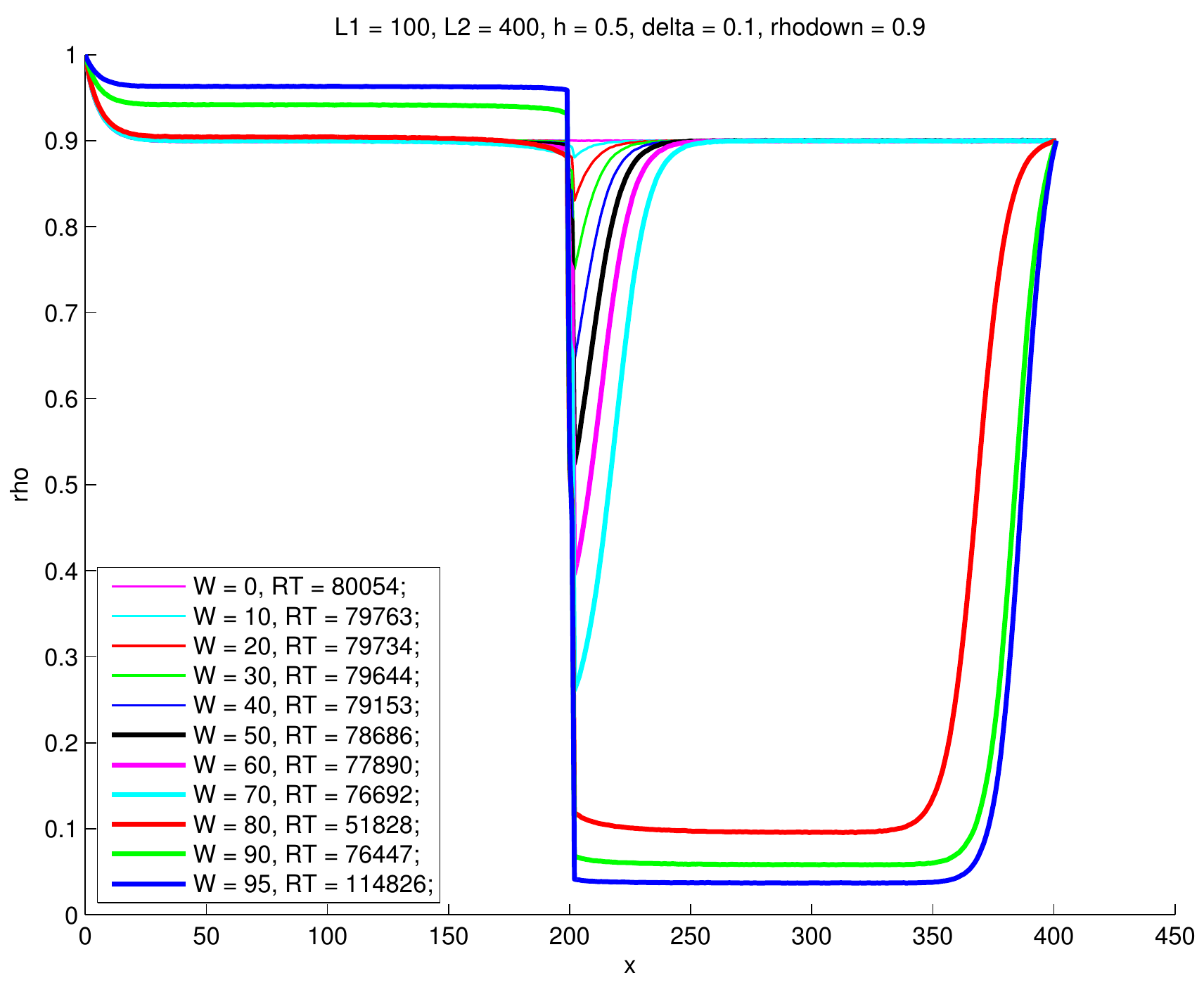}
    &
      \includegraphics[width=0.3\textwidth]{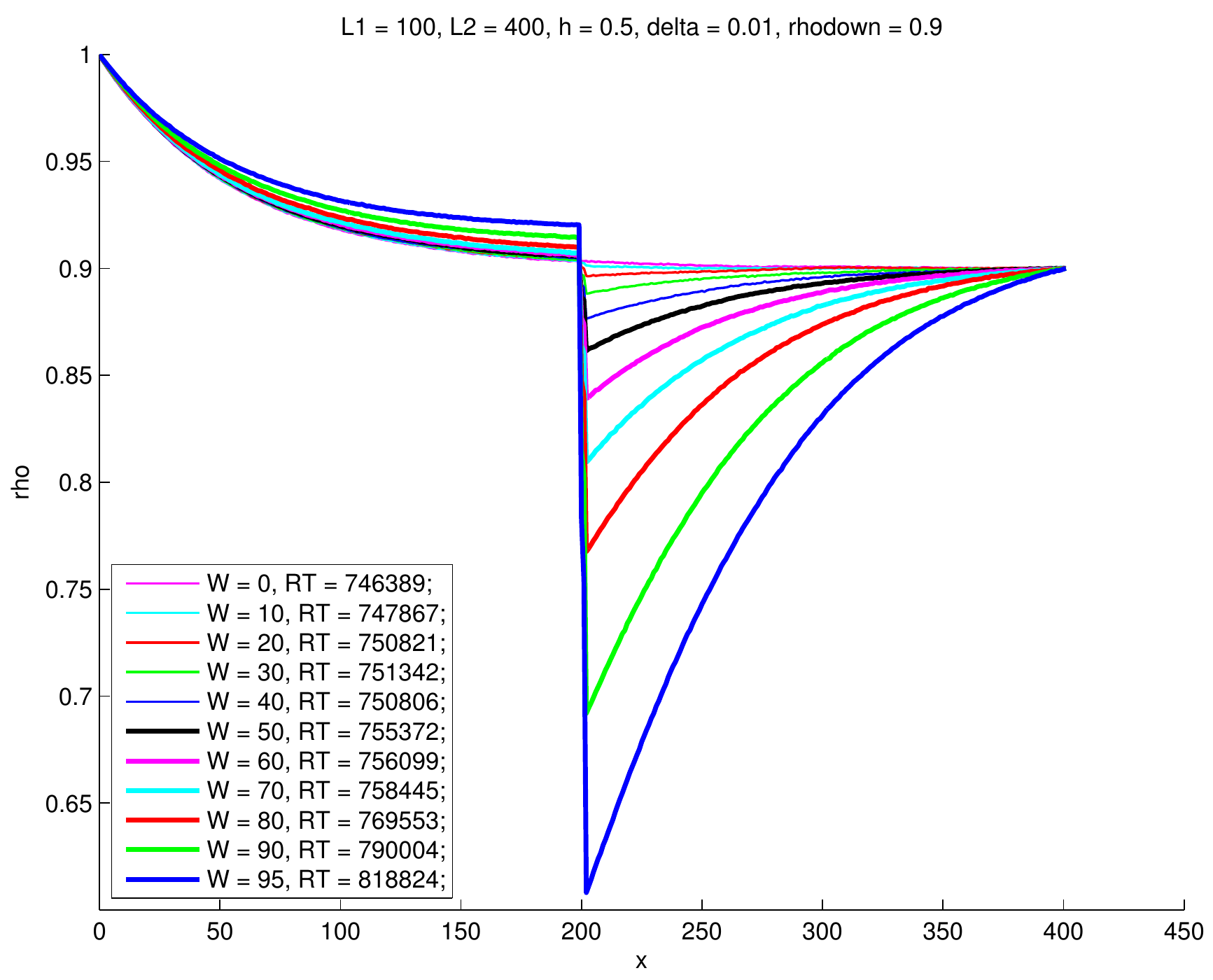}
  \end{tabular}
  \caption{The same as in Figure~\ref{fig:width-profile-comparison-drift-01-1}
  excepted for the bottom boundary density:
  $\rho_\rr{d}=0.7,0.8,0.9$ from the top to the bottom.
}
  \label{fig:width-profile-comparison-drift-01-2}
\end{figure*}

Figures~\ref{fig:width-profile-comparison-drift-01-1}
and \ref{fig:width-profile-comparison-drift-01-2}
show the profile $\tilde\rho(x)$ for different values of the parameters
$W=0,10,...,90,95$, $\drift=0.1$ and
$\drift=0.01$,
and bottom boundary density
$\rho_\rr{d}=0,0.4,0.5,0.6,0.7,0.8,0.9$. The remaining
parameters are not changed and are listed in the captions.

Here we see the drastic change in the density profile
behavior when the drift value decreases from $0.1$ to $0.01$.
When $\drift>0.1$ and $W=0$, the
density profile is nearly independent of $x$ and is equal to $0.5$.
It will not vary with $\rho_\rr{d}$, as long as $\rho_\rr{d}<0.5$. When
$\rho_\rr{d}>0.5$, the average value of the density increases with
$\rho_\rr{d}$ and is equal to it.
The case $\delta=0.01$ is reminiscent of the zero drift behavior, whereas
the case $\delta=0.1$ is qualitatively different.

We focus, now, in this latter case $\delta=0.1$.
Each panel in Figures~\ref{fig:width-profile-comparison-drift-01-1}
and \ref{fig:width-profile-comparison-drift-01-2}
refers to a fixed value of the bottom boundary density $\rho_\rr{d}$
and the different curves refer to different values of the
obstacle width $W$. In each a panel a quite obvious behavior
is observed: the jump in stationary density measured at the obstacle
increases with its width.

Much more interesting is the dependence of the density profile
on the bottom boundary density at fixed obstacle width.
Consider, for example, the case $W=50$ which corresponds to the black
lines in
Figures~\ref{fig:width-profile-comparison-drift-01-1}
and \ref{fig:width-profile-comparison-drift-01-2}.
The curve depicted in the top left panel in
Figure~\ref{fig:width-profile-comparison-drift-01-1}
refers to the case $\rho_\rr{d}=0$: in the upper part of the
strip (above the obstacle) the density profile is essentially
constant and drops to $0.73$ at the obstacle.
Immediately below the obstacle the density falls to $0.25$,
stays approximately constant for the whole bottom part
of the strip, and finally drops to zero to match
the boundary condition.

This behavior does not change that much when $\rho_\rr{d}$ is increased
till $\rho_\rr{d}=0.7$: the density profile lays, indeed, on the
reference
skeleton provided by the $\rho_\rr{d}=0$ case differing from it
only in the final part where the different boundary condition must
be matched. The picture is completely different when $\rho_\rr{d}$ gets
larger than the critical value $0.73$:
in the upper part of the strip the density profile
is approximately constant and equal to the bottom boundary
condition, whereas in bottom part it departs from the
reference skeleton and immediately below the obstacle it starts
increasing to match the boundary condition.

This description is qualitatively analogous for any value of the
obstacle width $W$. Obviously the value of the critical
density measured on the upper face of the obstacle in correspondence
of the skeleton profile obtained for $\rho_\rr{d}=0$ changes
with $W$, but it is interesting to remark that it tends
to $0.5$ (the value measure in absence of the obstacle)
as the width of the obstacle is decreased.

Summarizing, there exist
two different regimes (controlled by $\rho_\rr{d}$)
for $\delta>0$
of the obstacle, no onset of percolation.
When $\rho_\rr{d}>0.5$, depending on the obstacle width,
the system can be
in the low flux regime with onset of percolation.

This behavior is very similar to the phase transition which
is observed in the 1D simple exclusion model \cite{Krug}
with critical bottom boundary density $0.5$
(see, also, \cite{Rutger} for the
the obstacle free strip geometry \cite{Rutger}).
Here, the critical bottom density is not $0.5$ but it is given by
the density on the obstacle measured in the reference
skeleton profile corresponding to $\rho_\rr{d}=0$.
From the physical point of view the two phases differ
for the particle content in the bottom part of the strip:
such a region is almost empty in one case and pretty full in
the other. This behavior of the density profile has obviously
an important effect on the residence time.

As we have already remarked above, this qualitative change in
the density profile is not observed in the zero drift case.
Such a peculiar behavior that, as we shall see in the following,
has also a relevant consequence on residence times, is
due to the combined effects of the obstacle and the external drift.

\subsection{Residence time}
\label{rt-dr}
In all the cases discussed Section~\ref{dp-dr}
we have computed the residence time and reported it in the inset
in Figures~\ref{fig:width-profile-comparison-drift-01-1}
and \ref{fig:width-profile-comparison-drift-01-2}.
For convenience we summarize the data corresponding
to $\rho_\rr{d}=0,0.9$ in Tables~\ref{tab:compare_00}
and \ref{tab:compare_09} together with the Mean Field estimate
\eqref{rt-mf} computed by using the horizontally averaged
density profile $\tilde\rho(x)$.
We remark that in this case this procedure will not give
an accurate prediction for the residence time due to
the lacking of horizontal translational invariance in the density
profile, nevertheless, as the data will show, the prediction will
be at least qualitatively sound. Moreover, on the basis of the
Mean Field approximation it will possible to
interpret the Monte Carlo results.

\begin{table}[t]
  \centering
  \begin{tabular}{lllll|llll}
    $W$ & $R_{LA}$ & $R_{MF}$ & $m_1$ & $m_2$ & $R_{LA}$ & $R_{MF}$ & $m_1$ & $m_2$\\
    \hline
    0   & 15972    &   16165 & 80.5 & 200.9 & 124110 & 127267 &  633 & 201.1 \\
    10  & 16105    &   16215 & 80.7 & 200.9 & 124526 & 126405 &  629 & 200.9 \\
    20  & 16740    &   16577 & 82.5 & 201.0    & 125953 & 125501 &  624 & 201.0 \\
    30  & 18176    &   18476 & 92.0 & 201.0  & 128020 & 129696 &  645 & 201.0 \\
    40  & 20405    &   20906 & 104.0 & 200.9 & 131285 & 128462 &  639 & 201.0 \\
    50  & 23445    &   24252 & 120.6 & 201.0 & 135984 & 132939 &  662 & 200.9 \\
    60  & 27833    &   27914 & 138.9 & 201.0 & 142879 & 140895 &  701 & 201.0 \\
    70  & 34344    &   35402 & 176.1 & 201.0 & 152902 & 148000 &  736 & 201.1 \\
    80  & 45248    &   46583 & 231.8 & 201.0 & 169800 & 174161 &  866 & 201.2 \\
    90  & 69905    &   76254 & 379.4 & 201.0 & 205000 & 212743 & 1058 & 201.0 \\
    95  &103821    &  120039 & 597.3 & 201.0 & 252167 & 267808 & 1332 & 201.1 \\
  \end{tabular}
 \vskip 0.5 cm
  \caption{Comparison between the residence time of the lattice simulation with $\rho_\rr{d}=0.0$
    and its Mean Field approximation, based on the
averaged simulated density profile, along with its components.
The quantities $m_1$ and $m_2$ are defined in equation \eqref{prodotto}.
The other parameters are as in
Figure~\ref{fig:width-profile-comparison-drift-01-1}, in particular
$\drift=0.1$ is on the left and $\drift=0.01$ is on the right.}
  \label{tab:compare_00}
\end{table}

\begin{table}[t]
  \centering
  \begin{tabular}{lllll|llll}
    $W$ & $R_{LA}$ & $R_{MF}$ & $m_1$ & $m_2$ & $R_{LA}$ & $R_{MF}$ & $m_1$ & $m_2$\\
    \hline
    0  &  80054 &  80953 & 223.3 & 362.5 & 735757 & 803550 & 2186 & 367.6 \\
    10 &  79763 &  80066 & 221.0 & 362.2 & 737664 & 795020 & 2163 & 367.6 \\
    20 &  79734 &  82502 & 228.0 & 361.9 & 738747 & 810731 & 2207 & 367.4 \\
    30 &  79644 &  81053 & 224.5 & 361.1 & 739715 & 806241 & 2197 & 366.9 \\
    40 &  79153 &  81185 & 225.7 & 359.7 & 739925 & 812013 & 2217 & 366.3 \\
    50 &  78686 &  79679 & 222.8 & 357.7 & 743705 & 812770 & 2223 & 365.5 \\
    60 &  77890 &  79146 & 223.2 & 354.6 & 744789 & 831905 & 2283 & 364.4 \\
    70 &  76692 &  78093 & 223.4 & 349.5  & 746229 & 846801 & 2334 & 362.7 \\
    80 &  51828 &  54286 & 237.7 & 228.4 & 755008 & 858597 & 2382 & 360.5 \\
    90 &  76447 &  80147 & 369.3 & 217.0 & 773265 & 874445 & 2455 & 356.1 \\
    95 & 114826 & 120569 & 560.6 & 215.1 & 801341 & 877180 & 2501 & 350.8
  \end{tabular}
 \vskip 0.5 cm
  \caption{Comparison between the residence time of the lattice simulation with $\rho_\rr{d}=0.9$
    and its Mean Field approximation, based on the
averaged simulated density profile, along with its components.
The quantities $m_1$ and $m_2$ are defined in equation \eqref{prodotto}.
The other parameters are as in
Figure~\ref{fig:width-profile-comparison-drift-01-1}, in particular
$\drift=0.1$ is on the left and $\drift=0.01$ is on the right.}
  \label{tab:compare_09}
\end{table}

\begin{figure}[t]
  \centering
  \includegraphics[width=0.45\textwidth]{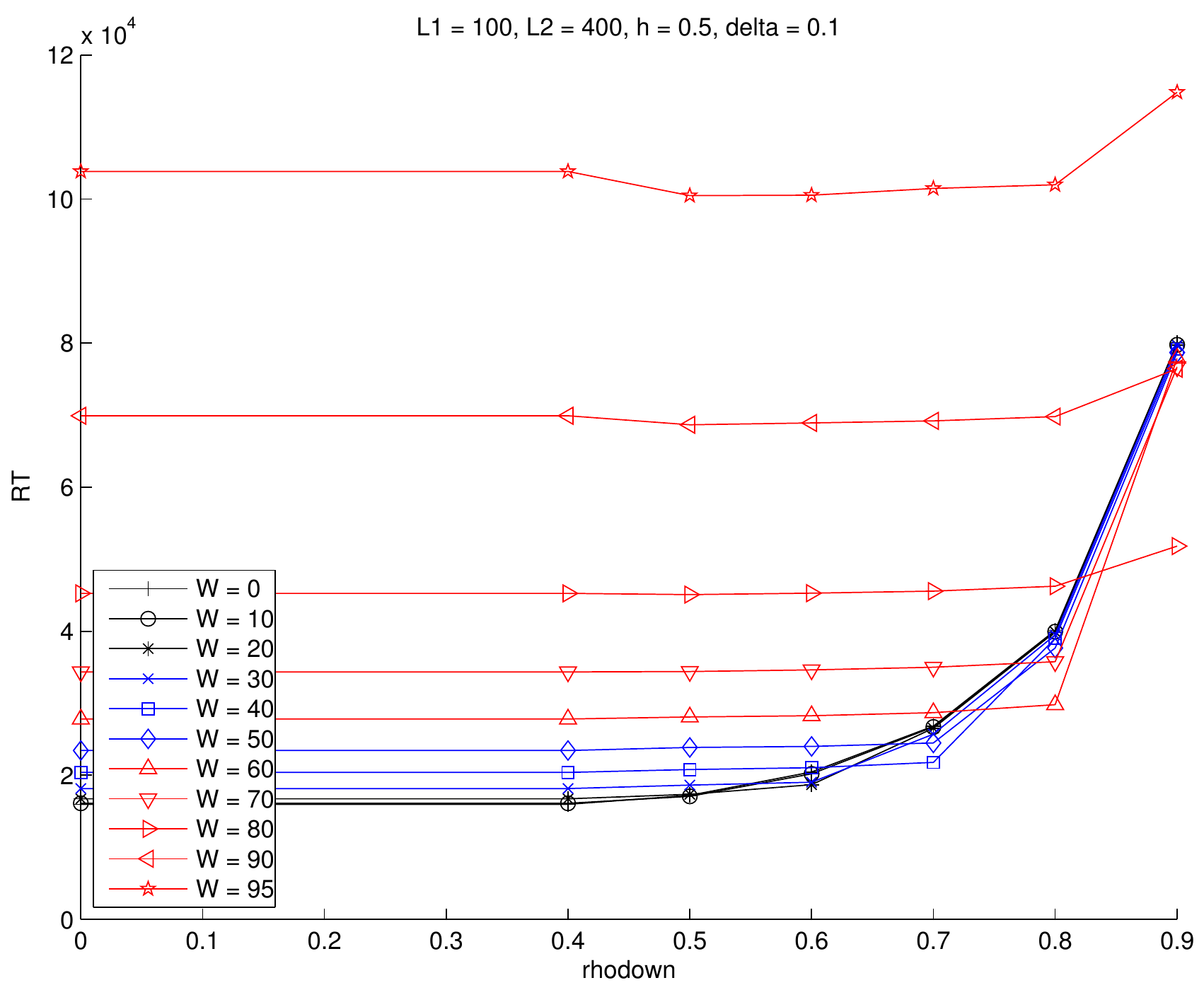}
  \caption{Residence time as a function of $\rho_\rr{d}$
for the different values of $W$ listed in the inset.
Parameters: $L_1=100$, $L_2=400$, $h=0.5$, $\drift=0.1$,
$\rho_\rr{u}=1$, and $O_2=3$.}
  \label{fig:rt_rhodown_comparison}
\end{figure}

In view of our results on the structure of the density
profile we know that, for $\delta>0$, there exists two different regimes
controlled by $\rho_\rr{d}$.
The difference in the residence time behavior is illustrated by
Figure~\ref{fig:rt_rhodown_comparison} that shows the residence time
as a function of $\rho_\rr{d}$.
Indeed, for
$\rho_\rr{d}<0.5$
there is only a weak dependence on $\rho_\rr{d}$, whatever is the
obstacle with $W$. For $\rho_\rr{d}>0.5$, depending on $W$,
there is a large increase of residence time with $\rho_\rr{d}$ itself.
A simple interpretation of this fact is the following: when the
bottom boundary density is large the system is in the low flux regime and
bottom part of the strip is so highly populated that the residence
time becomes large. But, as we shall see in the following, a deeper
understanding of this phenomenon can be achieved by means of the
Mean Field approximation.

Recall the Mean Field approximation \eqref{rt-mf}
of the residence time and note that $R$ is written as a product of two terms:
the area under the density profile and a factor depending on the
slope of the density profile at the top boundary.
Hence, it is convenient to introduce the quantities
\begin{equation}
\label{prodotto}
  m_1 = -\frac{2}{(1-h)\rho'(0)}
\;\;\textrm{ and }\;\;
  m_2 = \int_0^{L_2}\rho
\end{equation}
and write $R=m_1m_2$. The values of $m_1$ and $m_2$ for the
simulations discussed in this section are listed in
Tables~\ref{tab:compare_00} and \ref{tab:compare_09}.

A deeper insight in the problem is possible by looking at the
dependence of the residence time on $W$. We distinguish the two
regimes discussed above.

\emph{Regime $\rho_\rr{d}<0.5$.\/}
We note that the residence time increases uniformly with increasing $W$
as see in Figure~\ref{fig:rt_rhodown_comparison}.
Note also that the behavior of the residence time does not
depend on the value $\rho_\rr{d}$; we can then focus on the
case $\rho_\rr{d}=0$.
In terms of the Mean Field approximation this is due to
the variation of the derivative of the profile at the top boundary.
Indeed, see the values listed in the left part of
Table~\ref{tab:compare_00}, the parameter $m_2$ stays constant,
whereas $m_1$ steadily increases with the obstacle width.
This is due to an increased
density before the barrier that changes the slope $\rho'(0)$.
On the other hand,
the increase in density before the barrier and the developing wake
behind the barrier cancel and this explains why $m_2$ stays constant.

\begin{figure}[t]
  \centering
  \includegraphics[width=0.45\textwidth]{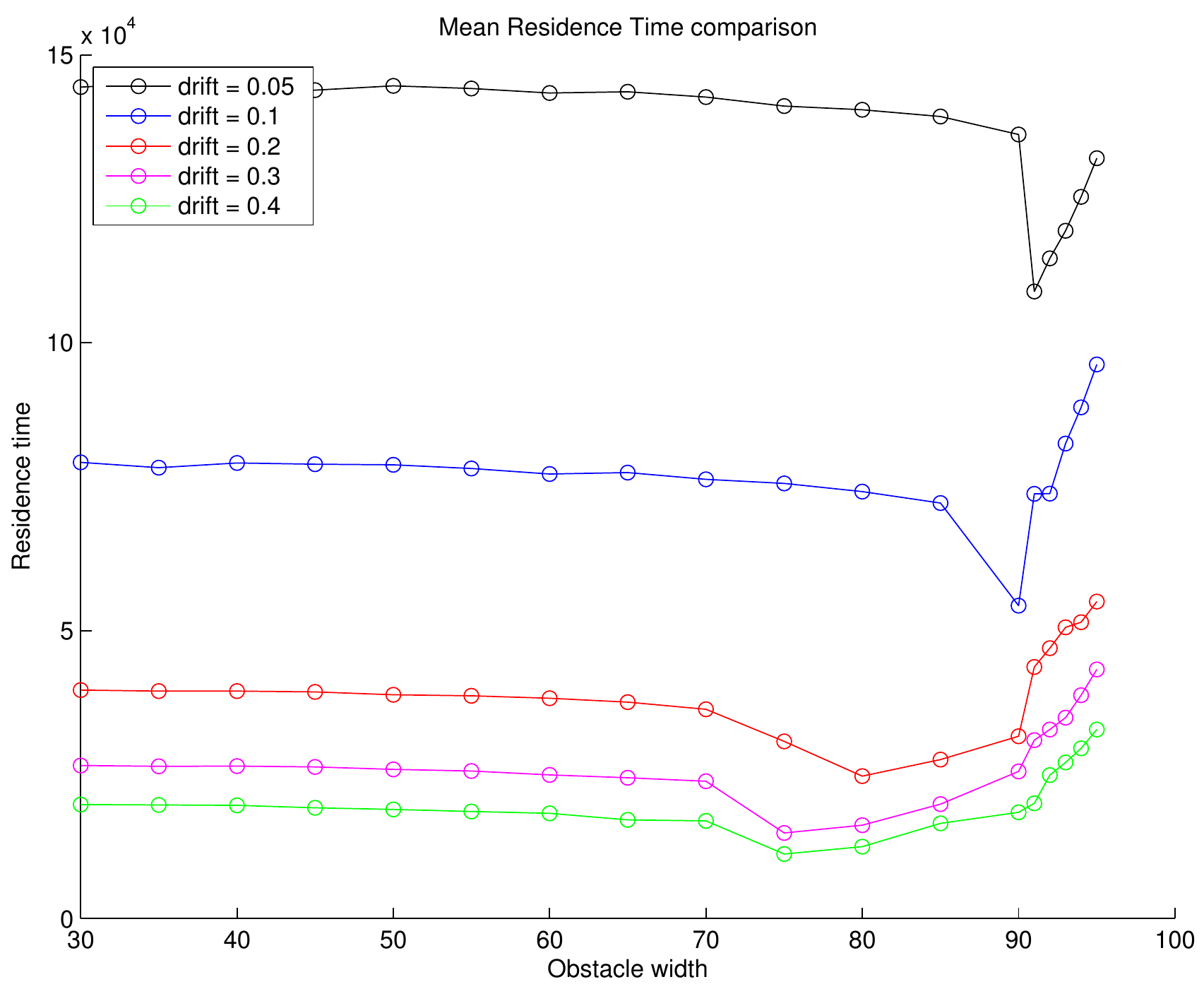}
  \caption{Residence time as a function of the obstacle width $W$
for the different values of $\delta$ listed in the inset.
Parameters: $L_1=100$, $L_2=400$, $h=0.5$,
$\rho_\rr{u}=1$, $\rho_\rr{d}=0.9$, and $O_2=3$.}
  \label{fig:nonlinear-rt}
\end{figure}


\emph{Regime $\rho_\rr{d}>0.5$.\/}
As long as the drift is small
(check the right value in the
Tables~\ref{tab:compare_00}-\ref{tab:compare_09}), the
  dependence of residence time on $W$ is similar to the case when the
  drift is zero. The dependence on $W$ is dominated by the diffusion
  and the residence time increases with increasing $W$, similar to the
  no drift case.

When $\drift$ is large, that is to say it
exceeds a particular value compared to the diffusion
  constant, a new dependence of the residence time on $W$ appears, see
  the data in the left in
  Table~\ref{tab:compare_09} and
  Figure~\ref{fig:nonlinear-rt}.
 Now there is an initial decrease in the
 residence time with increasing $W$ until a critical value of $W_\rr{c}$
  is reached where there's a dip in residence time.
  This critical width depend on the parameters of the model:
  from the full set of data listed in the insets in
  Figures~\ref{fig:width-profile-comparison-drift-01-1}
  and \ref{fig:width-profile-comparison-drift-01-2}
  it is possible to extract $W_\rr{c}$
  for all the considered values of
  $\rho_\rr{d}$ and observe that
  the ratio between the critical width $W_\rr{c}$ and $L_1$
  becomes close in value to $\rho_\rr{d}$ when $\rho_\rr{d}$ is high.
  Moreover, $W_\rr{c}$ decreases
  with drift and the width of the dip around $W_\rr{c}$ increases with
  drift.

This phenomenon is observed for any value of $\rho_\rr{d}$
larger than $0.5$, but the larger is $\rho_\rr{d}$ the more evident
the phenomenon is. We focus on the case $\rho_\rr{d}=0.9$ and $\delta=0.1$.
From the plots in the corresponding
panel on the left in
Figure~\ref{fig:width-profile-comparison-drift-01-2}
we see that,
when $W$ is increased from $0$ to $70$, the density profile in
the upper part of the strip remains essentially unchanged,
whereas a wake below the barrier appears.
The appearance of such a wake decreases the value of $m_2$ and hence
the residence time. Physically, it means the the number of particles
in the bottom part of the strip decreases and, thus, the typical time
to cross such a region gets lower.
This is confirmed by the data on the
left in Table~\ref{tab:compare_09}: the decrease of $m_2$ from
$362.5$ to $349.5$ causes the reduction of the residence time
from $80054$ to $75592$.

If $W$ is further increased, the coefficient $m_2$ goes
on decreasing, but the residence time
increases due to jamming. This is quantified from derivative and
density integrals. The system is in the fast flux regime,
so that the density increase in front of the barrier determines the
increase.  This is illustrated in Table~\ref{tab:compare_09},
indeed, the coefficient $m_1$, which is essentially constant for
$W=0,10,...,70$, starts to increase when $W$ exceeds $80$.

  We observe that initially there is no change in the density
  before the barrier (specifically in $\rho'(0)$), but a wake
  develops. The initial decrease in the residence time is therefore
  due to the decreased density of the wake. This dependence on $W$ is
  very different from the one observed in the previous cases, where
  there is always a substantial increase in density before the
  barrier. The difference must relate to the onset of percolation, so
  that a decrease in the wake density has a large effect.
  The distance dependence of the density profile
  in the wake changes from convex to concave beyond $W_\rr{c}$.
  The minimum in $W_\rr{c}$ occurs when the wake
  density reaches its minimum. Then there is no further reduction in
  the density possible and the density before the barrier now
  increases steeply with increasing $W$. The increase in residence
  time with $\rho_\rr{d}$ as well as the distribution before the barrier is
  now very similar to the other regime, that also suggests the
  importance of increased percolation, that results in jamming type
  behavior.



\begin{figure*}[t]
  \centering
  \begin{tabular}{llll}
    \includegraphics[width=0.20\textwidth]{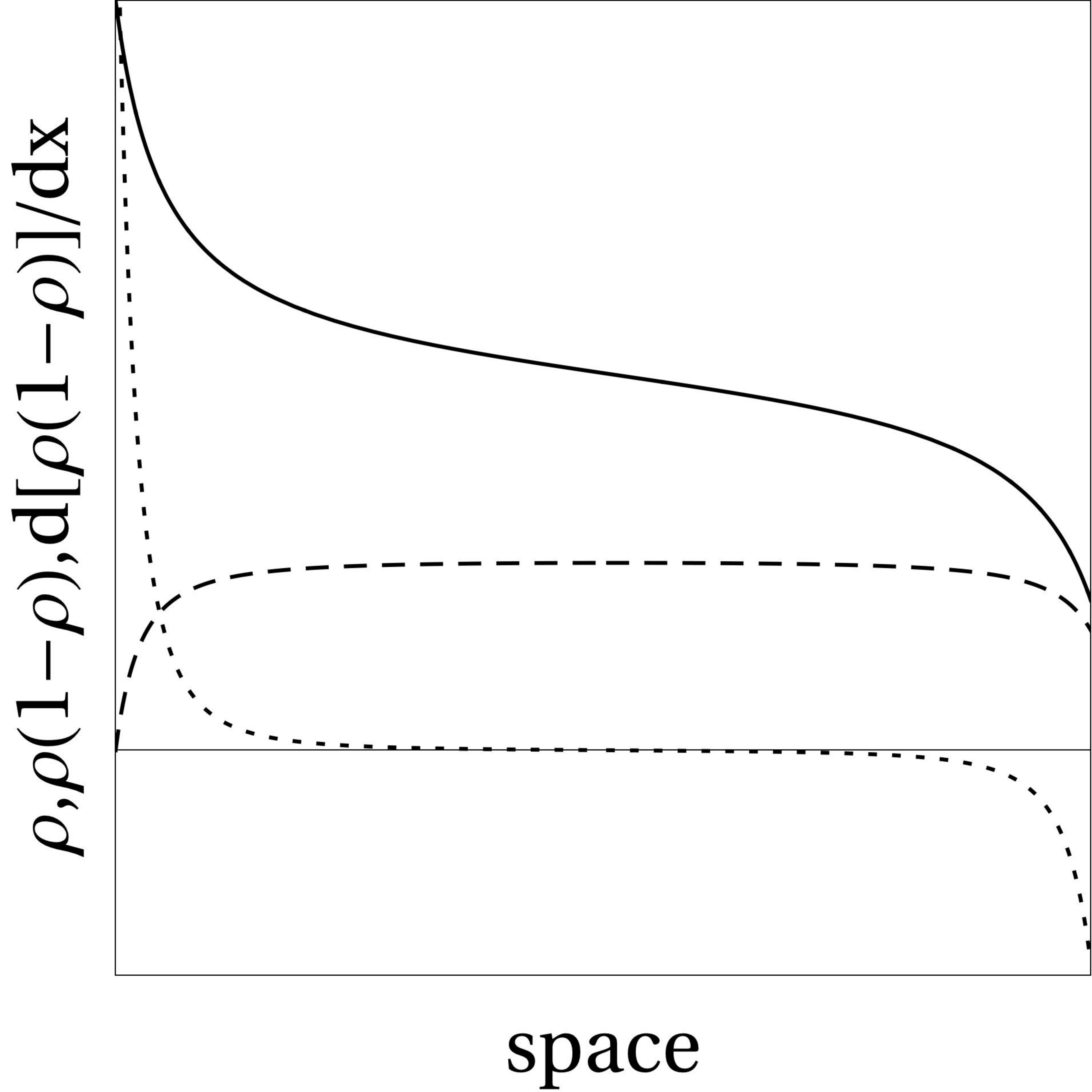}
    \quad\quad
    &
      \includegraphics[width=0.20\textwidth]{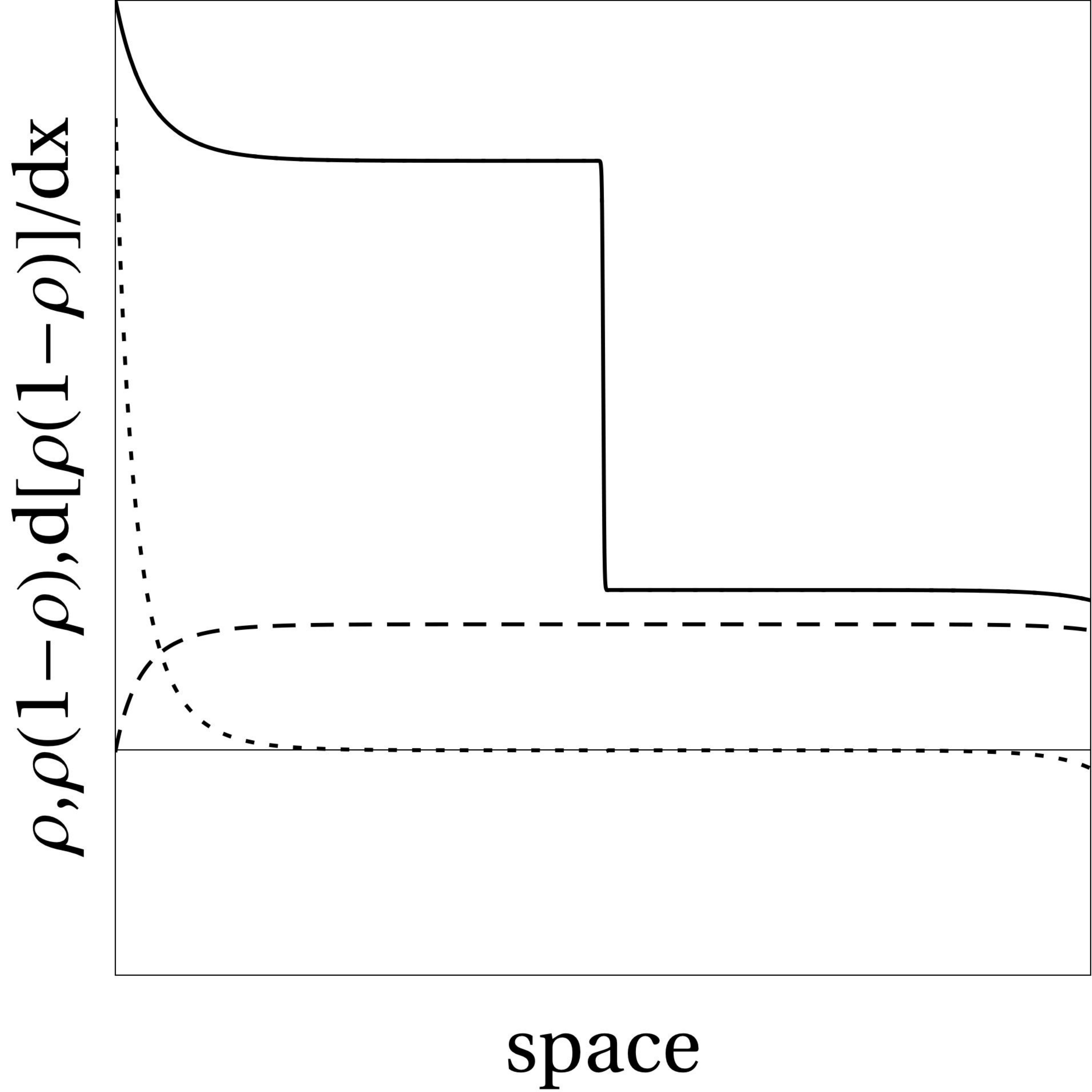}
      \quad\quad
    &
      \includegraphics[width=0.20\textwidth]{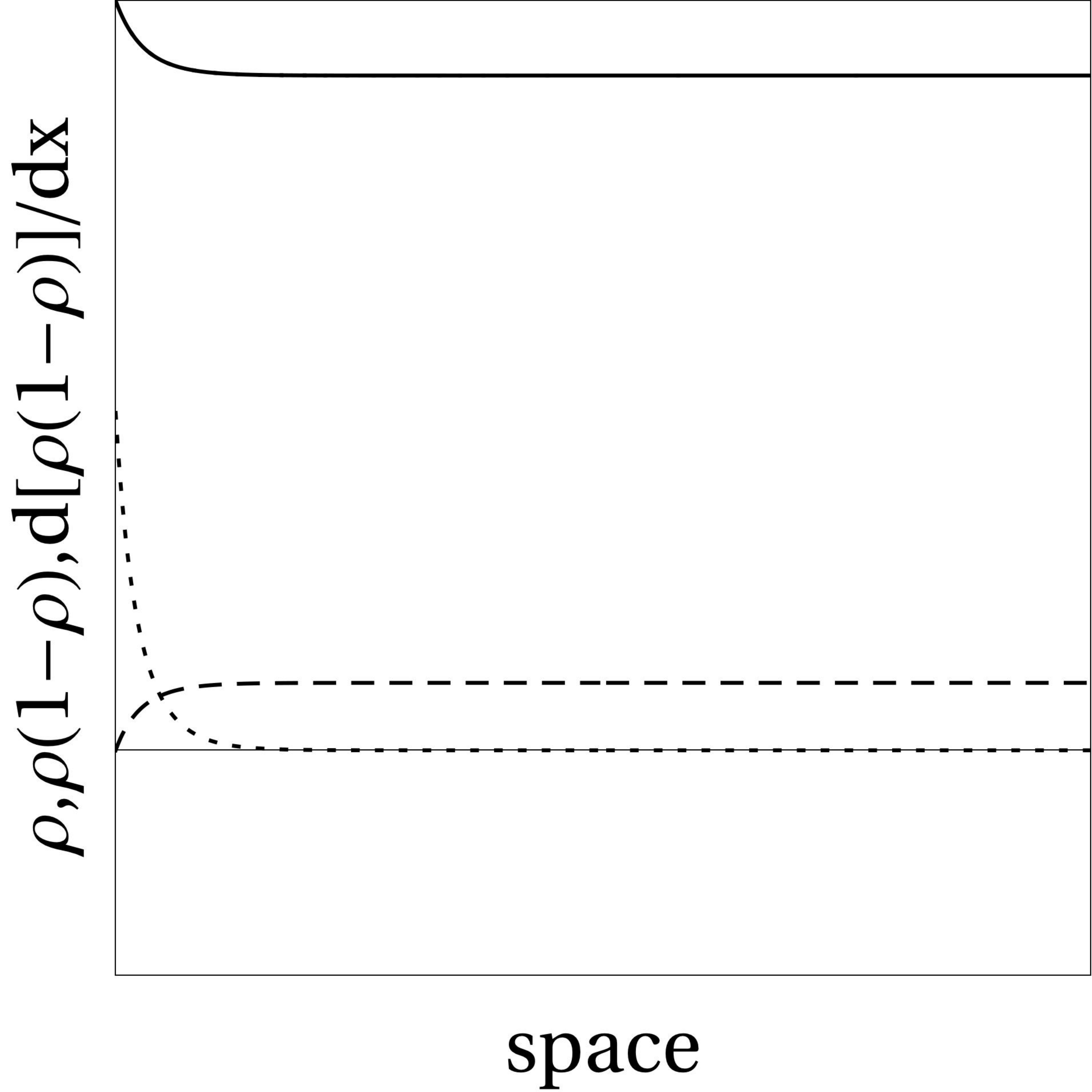}
      \quad\quad
    &
      \includegraphics[width=0.20\textwidth]{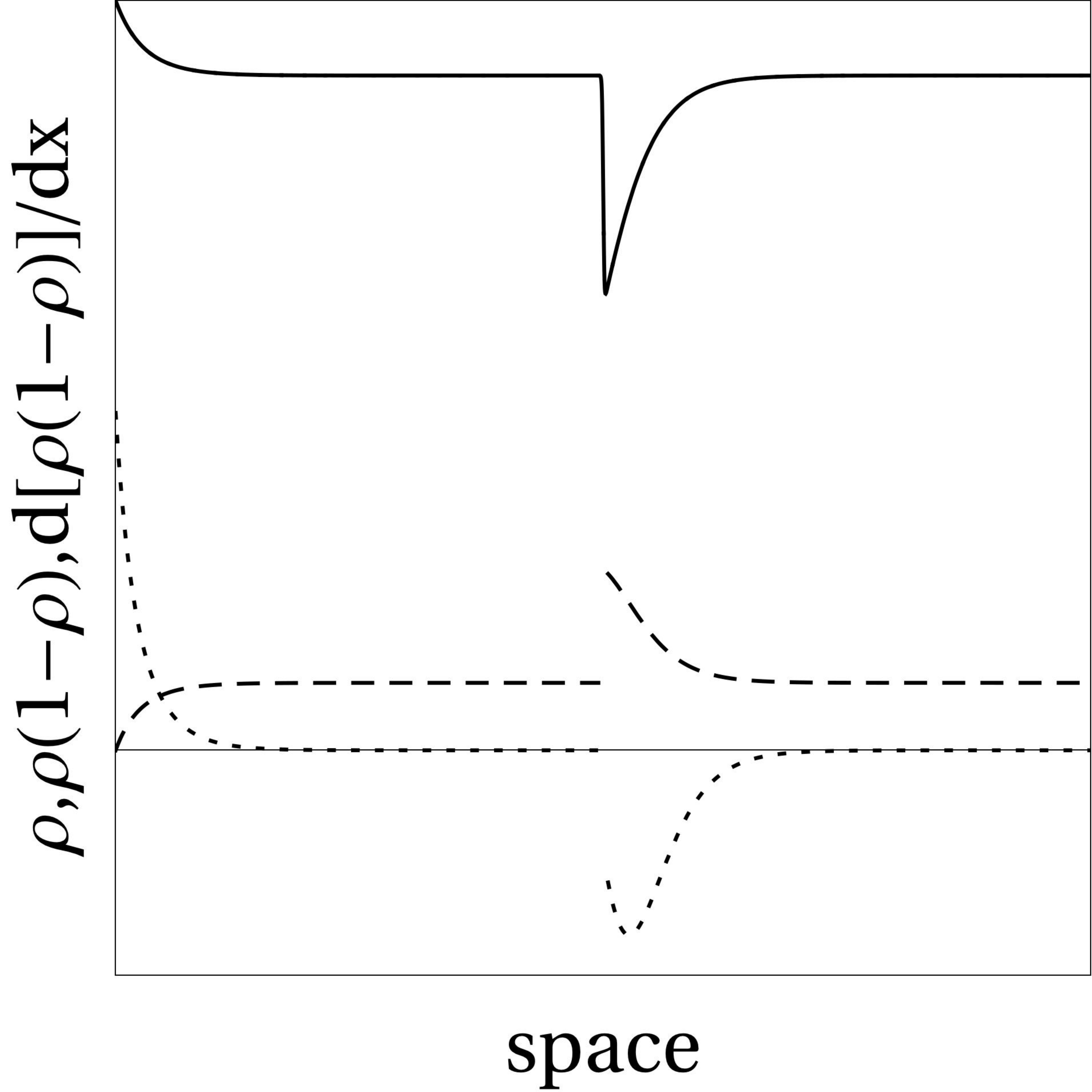}
    \\
  \end{tabular}
  \caption{Schematic (qualitative) representation of the
density profiles (solid lines) in the absence and presence of
barrier of small width and with not zero drift.
No quantitative information is provided in the picture except for the
zero which is represented by the thin solid horizontal line.
From the left to the right:
$\rho_\rr{d}< 0.5$ and no barrier,
$\rho_\rr{d}< 0.5$ with barrier at $L_1/2$,
$\rho_\rr{d}> 0.5$ and no barrier,
$\rho_\rr{d}> 0.5$ with barrier at $L_1/2$.
Dashed lines represent $\rho(x)[1-\rho(x)]$ and
dotted lines represent its derivative $\rr{d}\{\rho(x)[1-\rho(x)]\}/\rr{d}x$.
}
  \label{f:ultima}
\end{figure*}

\section{Discussion and conclusions}
\label{Discussion}
According to the lattice model simulations reported in this paper, the effect of
the barrier on the residence time is surprising: at
low flux the system may show decreased residence time of particles
when passage barrier is increased, instead of the expected decrease in
residence time.

We find three different flow regimes of interest.
The regime of zero drift, where the residence time increases
  with barrier length. The barrier generates an increase in density
  before the barrier and a wake behind. The density changes are
  comparable. The increase in residence time is due to the lowered
  derivative of density at the entrance of the stripe, that is due to
  the increased density before the barrier. The concentration of
  particles before the barrier is such that these particles can be
  considered to be in the percolation regime. The system becomes
  increasingly jammed.

The regime of non--zero drift, but with an exit density
  $\rho_\rr{d}<0.5$.  This is the regime of high flux
  (see \cite{Krug,Speer,Rutger}) without percolation inhibition. The
  residence time in this regime is independent of
  the bottom boundary
  density. When the barrier increases the residence time increases for
  a similar reason as in the zero--drift case.
  When $W/L_1$ becomes larger that $0.5$ the residence time
  increases steeply and the density before the
  barrier increases such that it is in the percolation regime. It is
  dominated again by the derivative of density at the entrance of
  strip.

The regime of non--zero drift, but with exit density
  $\rho_\rr{d}>0.5$.  This is the regime of low flux. When
  the Damk\"ohler number, i.e., the ratio between the external drift and
  the diffusion coefficient, is sufficiently large, the non--linear
  dependence of residence time on the barrier width appears.
  The residence time decreases with increasing value
  of barrier $W$ until a limiting value of $W_\rr{c}$ is reached.
  This critical width is such that its ratio with the
  horizontal length of the strip is
  equal or less than $\rho_\rr{d}$.
  Beyond this value of $W$ the
  residence time increases steeply with increasing $W$, as expected
  for the onset of increased jamming.

  The difference between the high
  flux and low flux regime is due to the very different dependence of
  $\rho(x)[1-\rho(x)]$ when $x\in [0,L_2]$.
  More precisely, consider the $\rho(x)$ plots in
  Figures~\ref{fig:width-profile-comparison-drift-01-1}
  and
  \ref{fig:width-profile-comparison-drift-01-2},
  the related graphs of the function $\rho(x)[1-\rho(x)]$
  behave as follows:
  as long as $\rho_\rr{d}<0.5$ the function $\rho(x)[1-\rho(x)]$ will have a
  maximum in $x\in [0,L]$,
  since $\rho$ varies between $\rho_\rr{d}\le 0.5$ and $1$.

The derivative of $\rho(x)[1-\rho(x)]$ varies from negative to positive
when $\rho'(x)>0$ or the opposite way when
$\rho'(x)\le0$.
When $\rho_\rr{d}>0.5$ the sign of the derivative of
$\rho(x)[1-\rho(x)]$ depends on $\rho'(x)$.
It is always  positive ($\rho'(x)<0$)
or negative  ($\rho'(x)>0$).
The contribution of drift to current
is proportional to the x derivative of
$\rho(x)[1-\rho(x)]$. Since this can only become zero,
when $\rho_\rr{d}<0.5$ or $\rho'(x)=0$,
the zero drift and drift curves only cross when $\rho_\rr{d}<0.5$.
This causes the density distribution when drift is not zero
to be less than that of the drift zero density distribution.
It becomes  equal to $\rho_\rr{d}$ over a large density regime when
$\rho_\rr{d}>0.5$ and drift exceeds a particular value
(this value falls between $0.1$ and $0.01$ in
Figures~\ref{fig:rt_rhodown_comparison}
and \ref{fig:nonlinear-rt}.

The changes in residence time
in presence of barrier can be understood as maximization of current.

In Figure~\ref{f:ultima} the changes in density for
$\rho_\rr{d}<0.5$ and $\rho_\rr{d}>0.5$ are schematically sketched.
There is an important qualitative difference between the two cases.

The case $\rho_\rr{d}<0.5$ (Figure~\ref{f:ultima} first and second panel
from the left):
the barrier reduces transmission from
$x<L_2/2$ to $x>L_2/2$,
since the density gradient at $x=L_2/2$ has the same sign before
and after barrier. Also directly behind the barrier
$\rho'(x)<0$, since this gives a positive contribution to
the flow rate,
the density dependence on $x$ then is concave.
Reduced transmission through the barrier increases the density before
the barrier into the $\rho_\rr{d}>0.5$ regime,
that is the percolation regime.
Behind the barrier a wake develops of lower density.
The flow rates before as well after the barrier decrease.

The case $\rho_\rr{d}>0.5$ (Figure~\ref{f:ultima} third and fourth panel
from the left):
when $\rho_\rr{d}>0.5$ the $x$ derivative of
$\rho(x)[1-\rho(x)]$ is negative as long as
$\rho'(x)>0$. An initial convex shape of density profile of
the wake behind the barrier implies $\rho'(x)>0$ ($x>L_2/2$).
The flow rate now increases, because in the wake the density is reduced
and $\rho(x)[1-\rho(x)]$ then increases. Density reduction when barrier
width is small is initially in the percolation regime.
Since the derivative $\rho'(x)$ before the barrier is negative and
positive after the barrier,
the second derivative of $\rho(x)$ is discontinuous.
Barrier transmission is not hindered as long as $W<W_\rr{c}$, where
the derivative $\rho'(x)$ changes sign.
When $W$ remains less than $W_\rr{c}$ there is no increase in density
before the barrier. At this condition the fast flow in the wake of barrier
drains density from the front of the barrier, so that it is maintained
at the density it also has in absence of the barrier.
The current increases with increasing barrier width, until no
density reduction in the wake is any more possible and the initial
sign of $\rho'(x)$ becomes negative. Then reduced transmission
through the barrier increases density before the barrier and current
decreases.

This analysis has been done for the projection of the two dimensional
changes in density onto a one dimensional density. In the two
dimensional case, a relative value of the horizontal displacement
$h=0.5$ has been used. In that case there is rapid diffusion of
density before the barrier to the opening positions between barrier
and wall, and after the barrier into the wake region. In the low flux
region,  the low density that develops in the wake also reduces density
between barrier and wall so that density transport from before the
barrier to the open space region is enhanced. The one dimensional
analysis indicates that asymmetrical density development is indeed
caused by the convex residence time function of $\rho$ in the high flux
region, that ultimately is due to percolation.

In the recent paper \cite{DMNS}
the totally asymmetric simple exclusion process has been applied to a
molecular motor transport model on a network. Whereas the network is
different from our strip model and drift equals one,
the paper \cite{DMNS} finds also non--linear dependence on motor
particle density when its global density exceeds a critical value
and network exit rate is asymmetric. Also in this case the critical behavior
depends on the derivative of $\rho(x)[1-\rho(x)]$ as we propose in this paper.

\begin{acknowledgments}
ENMC acknowledges ICMS (TU/e) for the kind hospitality and financial
support.
OK acknowledges support from the ``Over Grenzen, complexiteit programma"
of KNAW (Royal Netherlands Academy of Sciences and Arts).
\end{acknowledgments}

\end{document}